\documentclass[11pt]{article}
\usepackage{verbatim}
\usepackage{amsfonts,amsmath,amsthm}

\usepackage{latexsym}
\usepackage{amssymb}
\usepackage{amsmath}
\usepackage{graphicx}

\newtheorem{Theorem}{Theorem}[section]
\newtheorem{Definition}[Theorem]{Definition}
\newtheorem{Proposition}[Theorem]{Proposition}
\newtheorem{Assumption}[Theorem]{Assumption}
\newtheorem{Lemma}[Theorem]{Lemma}

\newtheorem{Remark}[Theorem]{Remark}

\usepackage{amssymb}
\usepackage{float}
\usepackage{epsfig}
\usepackage{amsmath}
\usepackage[english]{babel}
\usepackage{amstext}
\usepackage{mathrsfs}
\usepackage{amsthm}
\usepackage{fancyhdr}
\usepackage{latexsym}
\usepackage{amsmath}
\usepackage{amsfonts}
\usepackage{amssymb}
\usepackage{fancyhdr}

\usepackage[usenames]{color}
\definecolor{red}{rgb}{1.0,0.0,0.0}

\definecolor{blu}{rgb}{0.0,0.0,1.0}

\definecolor{green}{rgb}{0.0,0.7,0.0}

\def \N{\mathbb{N}}
\def \R{\mathbb{R}}

\def \E{\mathbb{E}}

\def \G{\mathbb{G}}

\def \P{\mathbb{P}}
\def \Q{\mathbb{Q}}

\def \Ac{{\cal A}}

\def \Ec{{\cal E}}
\def \Fc{{\cal F}}
\def \Gc{{\cal G}}
\def \Hc{{\cal H}}

\def \Sc{{\cal S}}

\def \Wc{{\cal W}}

\def \ni{\noindent}

\def \eps{\varepsilon}

\def \ep{\hbox{ }\hfill$\Box$}

\long\def\symbolfootnote[#1]#2{\begingroup%
\def\thefootnote{\fnsymbol{footnote}}\footnote[#1]{#2}\endgroup}

\def\reff#1{{\rm(\ref{#1})}}

\def\beqs{\begin{eqnarray*}}
\def\enqs{\end{eqnarray*}}
\def\beq{\begin{eqnarray}}
\def\enq{\end{eqnarray}}

\newcommand{\argmax}{\operatornamewithlimits{argmax}}

\date{}

\begin{document}

\date{\today}

\title{
Impact of time illiquidity in a mixed market without full observation\footnote{This research was partially supported by the PRIN research project ``Metodi deterministici e stocastici nello studio di problemi di evoluzione"  of the Italian Minister of University and Research.}}
\author{Salvatore Federico$^{\,a)}$  \and Paul Gassiat$^{\,b)}$ \and Fausto Gozzi$^{\,c)}$}
    \maketitle

\symbolfootnote[0]{$^{a)}$ Dipartimento di Economia, Management e Metodi Quantitativi, Universit\`a di Milano, Italy. E-mail: \texttt{salvatore.federico@unimi.it}.
Part of this research was done when this author was post-doc at  the LPMA - Universit\'e Paris 7. This author also acknowledges partial financial support of INdAM  (Istituto Nazionale di Alta Matematica).}
\symbolfootnote[0]{$^{b)}$ Institut f\"ur Mathematik, TU Berlin. E-mail: \texttt{gassiat@math.tu-berlin.de}.}
\symbolfootnote[0]{$^{c)}$ Dipartimento di Economia e Finanza, Libera Universit\`a degli Studi Sociali ``Guido Carli'', Roma, Italy. E-mail:
    \texttt{fgozzi@luiss.it}.}

\begin{abstract}
We study a problem of optimal investment/consumption over an infinite horizon in a market with two possibly correlated assets\;:  one liquid and one illiquid. The liquid asset is observed and can be traded continuously, while the illiquid one can be traded only at discrete random times, corresponding to the jumps of a Poisson process with intensity $\lambda$, is observed at the trading dates, and is partially observed between two different trading dates.
The problem is a nonstandard mixed discrete/continuous optimal control problem, which we solve by a dynamic programming approach. When the utility has a general form, we prove that the value function is the unique viscosity solution of the associated Hamilton-Jacobi-Bellman (HJB) equation and {characterize the optimal allocation in the illiquid asset}. In the case of power utility, we establish the regularity of the value function needed {to prove the} verification theorem, providing the complete theoretical solution of the problem.   This enables us to perform
numerical simulations, so as to analyze the impact of time illiquidity  and how this impact is affected by the degree of observation.
\end{abstract}

\noindent \textbf{Keywords:} Investment-consumption problem, liquidity risk,  optimal stochastic control, Hamilton-Jacobi-Bellman equation, viscosity solutions, regularity of viscosity solutions.

\bigskip

\noindent \textbf{MSC 2010 Classification}\,: 93E20, 91G80, 35D40, 35B65.\\
\medskip
\noindent \textbf{JEL  Classification}\,: C61, G11.


\section{Introduction}
Following the seminal works of Merton on portfolio management, a classical assumption in mathematical finance is to suppose that assets can be  continuously traded by the agents operating in the market. However, this assumption is unrealistic in practice, especially
in the case of less liquid markets, where investors cannot buy and sell assets
immediately, and have to wait  before being able to unwind a position.

In the recent years, several articles have studied the impact of this type of illiquidity. Rogers and Zane \cite{RZ02}, Matsumoto \cite{mat06}, Pham and Tankov \cite{phatan08} (see also \cite{CGPT,phatan09}) consider an investment model where the discrete trading times are given by the jump times of a Poisson process with constant intensity $\lambda>0$. Bayraktar and Ludkovski \cite{BL10} study a portfolio liquidation problem in a similar context.

The aforementioned papers focus on an agent investing exclusively in an illiquid asset. However, in practice it is common to have several correlated tradable assets with different liquidity. For instance, an index fund over
some given financial market will usually be more liquid than the individual tracked assets, while sharing a positive correlation with those assets. An investor in this market will then have the possibility of hedging his exposure in the less liquid assets by investing in the index and rebalancing his position frequently.

To our knowledge  few papers consider the case of a market with two (possibly correlated) assets, one liquid and one illiquid. This is the case of Longstaff \cite{L05}, who analyzes a two agents portfolio problem in  a market with a liquid asset and another asset that becomes non tradable  for a given time period.
Schwartz and Tebaldi \cite{ST04} consider a market with a liquid asset that can be traded continuously, and an illiquid asset that cannot be traded and is liquidated at a terminal date.
In a  recent paper, Ang, Papanikolaou and Westerfield \cite{APW}, in an infinite horizon framework with discounted power utility of consumption,  take a less restrictive point of view on the tradability of the illiquid asset, assuming, as in  \cite{CGPT,gasgozpha11,mat06,phatan08,RZ02}, that it may be traded at discrete random times.

\medskip
{
Following  \cite{APW,ST04},  we also consider a market with a liquid asset and an illiquid one.  In particular, as in \cite{APW}, the illiquid asset can be traded at some discrete random dates.
From the modeling side, the main novelty of our paper is that it treats the case of incomplete observation of the illiquid asset price between trading dates, modeled through  an observation parameter  interpolating the two extreme cases of full and no observation. This new feature  leads us to follow a different methodology than \cite{APW}, relying on the tool of viscosity solutions to study the associated HJB equation.}

 More precisely, we study a problem of optimal investment/consumption over an infinite horizon in a market consisting of a liquid and an illiquid asset. The liquid asset is continuously observed and can be continuously traded. The illiquid asset is correlated with the liquid one, with correlation parameter $\rho\in(-1,1)$, and  can be traded only at discrete random times, corresponding to the jumps of a Poisson process with intensity $\lambda>0$. We assume that the illiquid asset  can be observed at the trading dates (as in \cite{CGPT,gasgozpha11,mat06,phatan08}), but  introduce a new feature in the model - with respect to the aforementioned literature -  allowing  the possibility of  partial information between trading dates.
We introduce a parameter, $\gamma\in[0,1]$, measuring the degree of observation of the illiquid asset between two trading dates. The limit cases, $\gamma=0$ and $\gamma=1$, correspond, respectively, to the observation settings of  \cite{CGPT, phatan08,phatan09} and \cite{APW, RZ02, ST04}.

\medskip

The  mathematical problem is a nonstandard mixed discrete/continuous optimal control problem.
By means of a suitable use of Dynamic Programming, extending the idea of \cite{phatan08}, we show that the stochastic control problem between trading times can be written as an infinite horizon stochastic time-inhomogeneous control problem.
Then, we apply  the usual machinery of DP for such  problems and, using some results of \cite{FGtec},\footnote{In \cite{FGtec}, these results are proved for  $\gamma=0$. Their extension to the general case $\gamma\in[0,1]$ is straightforward, see Subsection \ref{sec:visc}.}  characterize the value function $\widehat{V}$ of this auxiliary problem as the unique viscosity solution of a HJB equation. At this stage, the viscosity characterization only pertains to  the optimal allocation in the illiquid asset.\footnote{{The outcome of this part of the analysis is already  in \cite{FGtec}, in the special  case $\gamma=0$.}} In order to go further   and characterize the optimal feedback allocation in the liquid asset  and the optimal feedback  consumption strategy, we need to prove a regularity result for $\widehat{V}$. This nonstandard regularity result and the related analysis of the optimal consumption and allocation in the liquid asset  were left out of the analysis in \cite{FGtec}. Here we  provide this result - Theorem \ref{prop:reg}, which is the main theoretical contribution of the paper -  in the special case of power utility.\footnote{Our assumption on the utility function covers only the case of positive power, unlike \cite{APW}. However the method can be modified to cover the case of negative power as well (see also Remark \ref{rem:negp}).} It   gives a full theoretical solution to the problem. A numerical scheme is proposed for implementation and numerical results are then provided and discussed  for different values of the relevant parameters $\gamma,\lambda,\rho$.

\medskip


Section 2 describes the market model and formulates the investment/consumption problem. Section 3 shows how, by a suitable dynamic programming principle, the problem can be reduced to a standard continuous time stochastic  control problem; it presents useful properties of the value functions - the original one and the auxiliary one - and   characterizes  them by means of viscosity solutions; finally,  it  characterizes  the optimal investment in the illiquid asset.  Section 4 solves  the problem in the case of power utility and provides an iterative scheme. Finally, Section 5 is devoted to the discussion of the numerical results obtained.

\section{Model and optimization problem}

\setcounter{equation}{0}

Consider a complete filtered probability space $(\Omega,\Fc,(\Fc_t)_{t\geq 0},\P)$ satisfying the usual conditions, on which are defined:
\begin{itemize}
\item[-]A Poisson process $(N_t)_{t\geq 0}$, with intensity $\lambda>0$. We denote  by
$(\tau_k)_{k\geq 1}$ its  jump times; moreover we set $\tau_0=0$.
\vspace{-.2cm}
\item[-] Two independent standard Brownian motions $(B_t)_{t\geq 0}$, $(W_t)_{t_\geq 0}$, independent also of the Poisson process $(N_t)_{t\geq 0}$.
\end{itemize}
\subsection{Market model}
The market model  consists of 
two risky assets with correlation $\rho\in(-1,1)$: \begin{itemize}
\item[-]
A liquid risky asset that can be traded continuously; given  $b_L, \sigma_L>0$, its price  $L_t$ evolves according to
\beq\label{LL}
dL_t & = & L_t\,( b_Ldt + \sigma_LdW_t).
\enq
\vspace{-1cm}
\item[-]An illiquid risky asset that can only be traded at the trading times $\tau_k$; given $b_I, \sigma_I>0$, its price $I_t$ evolves according to
\beq\label{II}
dI_t & = & I_t\,\Big(b_I dt + \sigma_I\, (\rho dW_t + \sqrt{1 - \rho^2} dB_t)\Big).
\enq
\end{itemize}
Without loss of generality, we assume $L_0 = I_0=1$. We also suppose the availability of  a riskless asset with deterministic dynamics. For simplicity, we assume that the interest rate on this asset is constant and equal to $0$.
\begin{Remark}
If the riskless  interest rate is not $0$, one needs  to add an extra term in all the equations. In the special case of power utility in Section \ref{Sec:power}, the assumption that the rate  is null is without loss of generality, as it can be eliminated through the discount factor  of the objective functional (the constant $\beta$ in \eqref{eq:optprob} below) by a suitable change of variables (see in \cite[p.\,189, Remark 2]{Kab}).
\end{Remark}
\subsection{Information}
The information setting is the following.
\vspace{-.2cm}
\begin{itemize}
\item[-]
The liquid asset $L$ is continuously observed.
\vspace{-.2cm}
\item[-] The illiquid asset $I$ is observed at the trading random times $(\tau_k)_{k\in\N}$.
\vspace{-.2cm}
\item[-] The illiquid asset $I$ is partially observed in the time interval $(\tau_k,\tau_{k+1})$.
\end{itemize}
To formalize the last  issue, we suppose
 that  the Brownian motion $B_t$  can be split as
 $$B_t\ =\ \gamma B^{(1)}_t+\sqrt{1-\gamma^2}B^{(2)}_t, \ \ \ \gamma\in[0,1],$$
where $B^{(1)},B^{(2)}$ are mutually independent  Brownian motions, also  independent of $W,N$, with $B^{(1)}$  observed and $B^{(2)}$  unobserved.
Let  $(\mathcal{N}_t)_{t\geq 0}$,   $(\mathcal{W}_t)_{t\geq 0}$, $(\mathcal{B}^{(1)}_t)_{t\geq 0}$ be the filtrations generated, respectively, by $N$, $W$, $B^{(1)}$.
Define the $\sigma$-algebra
$\mathcal{I}_t= \sigma\left(I_{\tau_k}\mathbf{1}_{\{\tau_k\leq t\}}, \ k\in\N\right)$,  $t\geq 0,$
and  the  filtration
$$\mathbb{G}^0\ :=\ (\Gc^0_t)_{t\geq 0}; \ \ \ \ \ \mathcal{G}^0_t\ = \ \mathcal{N}_t\vee \mathcal{I}_t\vee\mathcal{W}_t \vee\mathcal{B}^{(1)}_t \ =\  \sigma(\tau_k,I_{\tau_k}; \tau_k\leq t) \vee \mathcal{W}_t\vee \mathcal{B}^{(1)}_t.$$
The observation filtration  is
$\mathbb{G} = (\mathcal{G}_t)_{t\geq 0},$ where $\mathcal{G}_t  = \mathcal{G}^0_t\vee \sigma(\P\mbox{-null sets}).$
This means that, at time $t$, the agent has\,:
\begin{itemize}
\item[-] full information on the past of the liquid asset up to time $t$;
\vspace{-.3cm}
\item[-] full information on the trading dates of the illiquid asset realized before $t$ and on the price  of the illiquid asset at such trading dates;
\vspace{-.3cm}
\item[-] partial information (as described above) on  the price of the illiquid asset at $t$.
\end{itemize}
The parameter $\gamma$ measures how much information on $I$ is available in the random interval $(\tau_k,\tau_{k+1})$. The limit cases are:
\begin{itemize}
\item[-] $\gamma=0$, corresponding to having no information on  $B$   in the interval $(\tau_k,\tau_{k+1})$ (as in the information setting of  \cite{phatan09});
\vspace{-.2cm}
\item[-] $\gamma=1$, corresponding to  full information and recovering the information setting of \cite{APW}.
\end{itemize}

\begin{Remark}\label{rem:partialinfo}
In order to motivate this setting, note that the observation of  $L$ corresponds to the observation of the process $W$, while the  ``observation" of $B^{(1)}$ should be understood
as follows.
The price of the illiquid asset is observed  at $(\tau_k)_{k\in\N}$ (as in \cite{CGPT, phatan08, phatan09}), while, at different times $t\in(\tau_k,\tau_{k+1})$, the agent observes  $I^{(1)}_t$, evolving according to
$$dI^{(1)}_t\ =\ I^{(1)}_t\big(b_Idt+\sigma_I(\rho \, dW_t+\sqrt{1-\rho^2}\, \gamma \, dB^{(1)}_t)\big), \ \ \ \ \  I^{(1)}_{\tau_k}\ =\  I_{\tau_k}.$$
Then $I_t = I^{(1)}_t\cdot I^{(2)}_t,$ where
$dI^{(2)}_t = I^{(2)}_t\sigma_I\sqrt{1-\rho^2}\sqrt{1-\gamma^2}\,dB^{(2)}_t,$ $ I^{(2)}_{\tau_k}=1,$
is an unobserved component of $I$.
Between two trading dates, the price $I$ is partially known: the factor $I^{(1)}$ is observed, but $I^{(2)}$ is not.
{Within the interval $(\tau_k,\tau_{k+1})$ the knowledge of $(L,  I^{(1)})$ is equivalent to the knowledge of $(W, B^{(1)})$. In this sense $W $and $B^{(1)}$ are observed and the observation filtration is $\mathbb{G}$}.

\end{Remark}
\subsection{Trading/consumption strategies and wealth dynamics}
Define the set of admissible trading/consumption strategies as follows.
Consider all the triplets  $(c,\pi,\alpha)$ such that
\begin{itemize}
	\item[(h1)] $c=(c_t)_{t\geq 0}$ is a continuous-time nonnegative process, $(\Gc_t)_{t\geq 0}$-predictable, with  locally integrable trajectories; $c_t$ represents the consumption rate at time $t$;
	\vspace{-.2cm}
	\item[(h2)]$\pi = (\pi_t)_{t\geq 0}$ is a continuous-time process, $(\Gc_t)_{t\geq 0}$-predictable, with  locally square integrable trajectories; $\pi_t$ represents the amount of money invested in the liquid asset at time $t$;
	\vspace{-.2cm}
	\item[(h3)] $\alpha=(\alpha_k)_{k\in\mathbb{N}}$, is a discrete process, where $\alpha_k$ is  $\Gc_{\tau_k}$-measurable; $\alpha_k$ represents the amount of money invested in the illiquid asset in the interval  $(\tau_k,\tau_{k+1}]$.
\end{itemize}
Given an initial wealth $r\geq 0$ and a triplet  $(c,\pi,\alpha)$ satisfying (h1)--(h3), the wealth process $R$ is  obtained by recursion on $k\in\N$\,:
\beq
R_0 &=& r, \label{stateR1}\\
R_t &=& R_{\tau_k} + \int_{\tau_k}^{t} \big(\pi_s (b_L ds + \sigma_L dW_s)- c_s ds\big)  + \alpha_k \left(\frac{I_{t}}{I_{\tau_k}} -1\right), \ \ t\in(\tau_k,\tau_{k+1}].\label{stateR2}
\enq
In general $R$ is  not $\mathbb{G}$-predictable (unless $\gamma=1$), as $I$ is not.
Following \cite[Sec.\,3, p.\,9]{APW} and \cite[Sec.\,2, p.\,7]{ST04}, we split  $R$ into:
\begin{itemize}
\item [-]  a liquid part $X$  (observable), containing  the money held in the liquid asset, the money held in the bank account  and the consumption;
\vspace{-.2cm}
\item[-] an illiquid part $A_t$  (partially observable).
\end{itemize}
They are defined in the intervals $[\tau_k,\tau_{k+1})$, $k\in\N$, as
\beq\label{def:X}
X_t&=&R_{\tau_k} -\alpha_k+  \int_{\tau_k}^{t} \big(\pi_s (b_L ds + \sigma_L dW_s)- c_s ds\big),
\enq
\beq \label{def:Z}
A_t&=&\alpha_k \,\frac{I_{t}}{I_{\tau_k}}.
\enq
Obviously
\beq\label{def:R}
R_t&=&X_t+A_t, \ \ \ \ \forall t\geq 0.
\enq
Observe that the process $R$ is continuous, while the processes $X,A$ are not, due to the rebalancing. Moreover, at time $\tau_k$ the process $R$ does not depend on the value of $\alpha_k$, whereas the processes $X,A$ do.

The class of admissible controls is the set of triplets  of processes $(c,\pi,\alpha)$ satisfying (h1)--(h3) and such that the corresponding wealth process $R$ is nonnegative (no-bankruptcy constraint). The latter class depends on the initial wealth $r$. Denote it by $\mathcal{A}(r)$ and note that it is not empty for every $r\geq 0$, as the null strategy $(c,\pi,\alpha)=(0,0,0)$ belongs to it.
As $\rho\in(-1,1)$, the illiquid asset  may become very large or  small, independently of what happens to the liquid asset.
Hence, having a short position in the illiquid asset or having a negative liquid wealth implies a positive probability of negative wealth.
 These facts suggest that requiring the positivity of $R$ should be equivalent to requiring the positivity of both $X$ and $A$.
\begin{Proposition}\label{prop:admiss} Let $r\geq 0$. The following facts are equivalent:
\begin{enumerate}
\item
$(c,\pi,\alpha)\in\mathcal{A}(r)$;
\item
$X_t\geq 0$, $A_t\geq 0$, for every $t\geq 0$;
\item $(c,\pi,\alpha)$ fulfills (h1)--(h3),
 $
0  \leq \alpha_k  \leq  R_{\tau_k}$ for every $k\in\N$, and
$$
- \int_{\tau_k}^{t} \big(\pi_s (b_L ds + \sigma_L dW_s)- c_s ds\big)  \ \leq\  R_{\tau_k} - \alpha_k,\ \ \ \ \forall t \in[\tau_k, \tau_{k+1}), \ \forall k\in\N.$$
\end{enumerate}
\end{Proposition}
\textbf{Proof.}
$3 \Leftrightarrow 2 \Rightarrow 1$ is straightforward, so it only remains to prove $1 \Rightarrow 2$.
Fix $(c,\pi,\alpha)\in\mathcal{A}(r)$ and  $t\geq 0$. Let  $s>t$,  $k\in\N$, and consider the non-negligible event
 $E_{s,k} := \{ \tau_k \leq t < s < \tau_{k+1}\}$ and the probability $\mathbb{P}_{s,k}(\cdot)=\frac{\mathbb{P}(\cdot\cap E_{s,k})}{\mathbb{P}(E_{s,k})}$. As $\cup_{s>t,\,k\in\N}E_{s,k}=\Omega$, it suffices to show that, for each $s>t$ and $ k\in\N$, we have
$ X_t \geq 0$ and  $A_t \geq 0$,  $\P_{s,k}-\mbox{a.s.}$.
So we work on the probability space $(E_{s,k},\  \mathcal{F}\cap E_{s,k},\ \mathbb{P}_{s,k})$ and consider, in the interval $[t,s]$, the filtration
$\mathbb{H}:=(\mathcal{H}_u)_{u\in[t,s]}$, with $\mathcal{H}_u := \mathcal{G}_u \vee \sigma(B_r ; \ r \geq 0),$
where, with an abuse of notation, we still indicate by $\mathcal{G}_u$ the $\sigma$-algebra $\mathcal{G}_u$ restricted to $E_{s,k}.$
The idea behind the use of the filtration $\mathbb{H}$ is that,  as the fluctuations of $I$ due to $B$ cannot be hedged, the agent who wants to check at time $t$ the admissibility  of a strategy has to take into account all the possible scenarios of $B$. So, conditioning the future wealth with respect to (the present information $\mathcal{G}_t$ and) $B$, the agent must get an almost surely nonnegative random variable.
In the rest of the proof, all the equalities and inequalities are intended $\P_{s,k}$-a.s..

As $B$, $W$, $N$ are independent, $W$ is still a Brownian motion under this filtration in the probability space defined above. By a Girsanov change of measure, if necessary,  without loss of generality we can take $b_L=0$. Then, letting
$T_n := \inf\left\{u\in[t,s] \, |  \, \int_t^{u} \pi_r \sigma_L dW_r \leq - {n} \right\}$, the process   $\left(\int_t^{u \wedge T_n} \pi_r \sigma_L dW_r -c_rdr\right)_{t\leq u \leq s}$ is a $\mathbb{H}$-supermartingale, thus from \eqref{def:X} it follows $\E_{s,k} [X_{s \wedge T_n} | \mathcal{H}_t]\leq X_t$. Hence,  from \eqref{def:Z}-\eqref{def:R}, conditioning with respect to $\mathcal{H}_t$,
\beqs
\E_{s,k} \left[R_{s \wedge T_n}\ \big|\  \mathcal{H}_t \right] &= &\E_{s,k} \left[X_{s \wedge T_n} + A_{s \wedge T_n}\ \big|\ \mathcal{H}_t \right]\ \leq \  X_t + \alpha_k \frac{\E_{s,k} \left[I_{s \wedge T_n}\ \big|\ \mathcal{H}_t \right]}{I_{\tau_k}}.
\enqs
Letting $n \to \infty$, we can  apply Fatou's lemma on the left hand side (by assumption $R\geq 0$) and dominated convergence on the right hand side, obtaining
\beqs
\E_{s,k} \left[R_{s}\, \big|\, \mathcal{H}_t \right] &\leq& X_t + \alpha_k \frac{\E_{s,k} \left[I_{s}\, \big|\, \mathcal{H}_t \right]}{I_{\tau_k}}.
\enqs
As $(c,\pi,\alpha)\in\mathcal{A}(r)$, we obtain
\beq\label{formula}
0&\leq& X_t + \alpha_k \frac{\E_{s,k} \left[I_{s} \,\big|\, \mathcal{H}_t \right]}{I_{\tau_k}}.
\enq
Let us  exploit this inequality by looking at the conditional law of its right hand side given $\mathcal{G}_t$. On $E_{s,k}$ we can decompose $I_s = I^{(1)}_s I^{(2)}_s I^{(3)}_s,$ where
$$
I^{(1)}_s=e^{(b_I-\sigma_I^2/2)s+\sigma_I (\rho W_t+\sqrt{1-\rho ^2} B_t)},  \ \  I^{(2)}_s=e^{\sigma_I \sqrt{1-\rho^2} (B_s-B_t)}, \ \ \ I^{(3)}_s= e^{\sigma_I\rho (W_s-W_t)}.
$$
$I^{(i)}_{s}$, for $i=1,2,3$, are lognormal and, resp., $\mathcal{G}_t$-measurable, $\sigma(B_u-B_t,\  u\in[t,s])$-measurable, and $\sigma(W_u-W_t, \ t\leq u\leq s)$-measurable. As $W_\cdot-W_t$ is independent of $\mathcal{H}_t$, we have  $\E_{s,k} \left[I_{s}\, \big|\, \mathcal{H}_t \right] =  I^{(1)}_s I^{(2)}_s \E[I^{(3)}_s]$ and \eqref{formula} becomes
\beq\label{formula1}
0&\leq& X_t + \frac{\alpha_k}{I_{\tau_k}} \,\E[I^{(3)}_s] I^{(1)}_s{I}^{(2)}_{s}.
\enq
Note that $I^{(1)}_s, X_t, \alpha_k, I_{\tau_k}$ are $\mathcal{G}_t$-measurable. On the other hand, $I^{(2)}_s$ is independent of $\mathcal{G}_t$, hence the conditional  law of $I^{(2)}_s$ given $\mathcal{G}_t$ is lognormal and nondegenerate, as $|\rho|<1$. In particular,  it  has full support in $(0,\infty)$.
Then, taking into account that $I_{\tau_k}>0$, $I^{(1)}_s>0$, $\E[I^{(3)}]>0$, it is clear that, to have \eqref{formula1},  it must be $X_t \geq 0$  and $\alpha_k\geq 0$. The latter is equivalent to $A_t \geq 0$   and we conclude.
\hfill$\square$
\subsection{Optimization problem}
The optimization problem consists in maximizing, over the set  $\mathcal{A}(r)$, the expected discounted utility of consumption over an infinite horizon\,: given a utility function $U$ and a discount factor $\beta>0$, the optimization problem is
\beq\label{eq:optprob}
\mbox{Maximize} \ \ \ \E \left[ \int_0^{\infty} e^{-\beta s} U(c_s) ds \right], \ \ \ \mbox{over} \  {(c,\pi,\alpha) \in \Ac(r)}.
\enq

%
\begin{Assumption}\label{ass:U}
The preferences of the agent are described by a utility function $U: \R_+\rightarrow \R$ continuous,  nondecreasing, concave, such that $U(0)=0$, and  the following growth condition holds\,: 
there exist $K_U>0$,  $p\in(0,1)$ such that
$
U(c)\leq K_U\,\frac{c^p}{p}.
$
\end{Assumption}

\begin{Remark}\label{rem:negp}
In applications one is often  interested in power utility functions
$U(c) = \frac{c^p}{p}$, $p\in (-\infty,1),$
with the convention that $U(c)=\log c$ when $p=0$. Assumption \ref{ass:U}
 includes only the case $p \in (0,1)$. The case of negative exponent
is interesting as well, as it seems to capture agents' behavior (see \cite{BW96}). We work with Assumption \ref{ass:U}, but  stress that the case $p\leq 0$ can be treated {similarly by suitable modifications, even if  a bit more difficult to handle (see also Remark 2.6 in \cite{CGPT})}. It  is treated in \cite{APW} under full observation.
\end{Remark}
\begin{Assumption}\label{ass:beta}
The discount factor $\beta$ is such that $\beta > k_{p}$, where
%
%
\beq\label{kp}
k_{p} := \sup_{u_L\in \R ,u_I \in [0,1]} \left\{p (u_L b_L + u_I b_I) - \frac{p(1-p)}{2} ( u_L^2 \sigma_L^2 + u_I^2 \sigma_I^2 + 2 \rho u_L u_I \sigma_L \sigma_I)\right\}.
\enq
\end{Assumption}

\begin{Remark}\label{remMerton}
{
The assumption on $\beta$ is related to the investment/consumption problem in a liquid market.
Let $p\in(0,1)$ and consider an agent with initial wealth $r$, consuming at rate $c_t$, investing in $L_t$, $I_t$ continuously, with respective proportions $u^L_t$, $u^I_t$, and under the constraint that $u^I_t \in [0,1]$. Suppose that preferences are represented by the utility function $U^{(p)}(c)=c^p/p$, with $p\in (0,1)$. Denote by $\Ac_{{Mert}}(r)$ the set of strategies keeping wealth nonnegative and define
\beq
V^{(p)}_{Mert}(r) &:=& \sup_{(u^L,u^I,c) \in \Ac_{Mert}(r)} \E \left[\int_0^\infty e^{-\beta t} U^{(p)}(c_t) dt\right], \label{defVM}
\enq
 This is a constrained Merton problem which dominates our problem, in the sense that $V^{(p)}_{Mert}(r)$ is higher than  the optimal value of our problem, up to the multiplicative constant $K_U$ of Assumption \ref{ass:U}.  One can see (for instance  solving the HJB equation) that $V^{(p)}_{Mert}$ is finite if and only if Assumption \ref{ass:beta} is satisfied
 and that, in this case,
$V^{(p)}_{Mert}(r) =  \left(\frac{1-p}{\beta - k_{p}}\right)^{1-p} r^p.
$
Therefore, Assumption \ref{ass:beta} guarantees, together with the growth condition of Assumption \ref{ass:U}, finiteness for our problem too.

Further note that the constrained liquid investment/consumption problem described above can always be reduced to the case where the two assets are independent, because
\beqs
dX_t\ \ = \ \ X_t\Big(u^L_t \frac{dL_t}{L_t} + u^I_t \frac{dI_t}{I_t}\Big) &=&X_t \left(\Big(u^L_t +  \frac{\rho b_L \sigma_I}{\sigma_L} u^I_t\Big) \frac{dL_t}{L_t} + u^I_t \frac{dJ_t}{J_t}\right),
\enqs\\
where $J$ is the process defined below in \reff{XYt} (taking $\gamma$ $=$ $0$),
and the problem is equivalent to an agent investing in $L$ and $J$, with the same constraint for the proportion invested in $I$.
However, this reduction does not work for the illiquid problem that we consider: neither the observation constraint (the integrand in $L$ being $\G$-adapted), nor the trading constraint (the amount held in the illiquid asset being constant between $\tau_k$ and $\tau_{k+1}$) are preserved by this transformation.
}
\end{Remark}

From now on Assumptions \ref{ass:U} and \ref{ass:beta} will be standing assumptions.

\section{Dynamic Programming}
\setcounter{equation}{0}
We denote the value function of the optimal stochastic control problem \eqref{eq:optprob} by $V$:
\beq\label{DPP1}
V(r)& := &\sup_{(c,\pi,\alpha) \in \Ac(r)} \E \left[ \int_0^{\infty} e^{-\beta s} U(c_s) ds \right], \ \ \ r\geq 0.
\enq
\begin{Proposition}\label{prop:V}
$V$ is everywhere finite, concave, $p$-H\"older continuous  and nondecreasing.  Moreover, there exists  $K_V>0 $ such that
\beq\label{estimV}
{V}(r)&\leq & K_V r^p, \ \ \ \ r\geq 0.
\enq
\end{Proposition}
{\bf Proof.} As observed in Remark \ref{remMerton}, finiteness and \eqref{estimV} follow from the growth condition of Assumption \ref{ass:U} and Assumption \ref{ass:beta}, by comparing with a constrained Merton problem.
Concavity of $V$ follows, by standards arguments, from concavity of $U$ and linearity of the state equation. Monotonicity follows, by standard arguments, from  monotonicity of $U$. Finally, $p$-H\"older continuity follows from concavity, monotonicity of $V$, and  \eqref{estimV}. \hfill$\square$\\\\
Following \cite{phatan08}, we state a  Dynamic Programming Principle (DPP) to reduce our mixed discrete/continuous problem to a standard one between two trading times.

\begin{Proposition}[DPP] \label{propDPP}
We have
\beq \label{eqDPP0}
V(r) &=& \sup_{(c,\pi,\alpha)\in \Ac(r)} \E\left[ \int_0^{\tau_1} e^{-\beta s} U(c_s) ds + e^{-\beta \tau_1} V\left(R_{\tau_1} \right) \right], \ \ \ \forall\, r\geq 0.
\enq
\end{Proposition}

{\bf Proof.}
{The proof parallels \cite{phatan09}. We only indicate modifications. The main difference is that, in our setting, there is some additional information between $\tau_n$ and $\tau_{n+1}$ brought by ${(\Wc_\cdot\vee {\cal B}^{(1)}_\cdot})$, so that the processes are no longer {deterministic on $(\tau_n,\tau_{n+1}]$  given $\mathcal{G}_{\tau_n}$, but only predictable with respect to $(\mathcal{W}_\cdot\vee {\cal B}^{(1)}_\cdot)$}. Then, one has to use the fact that a process $(\xi_t)_{t \geq 0}$ is $\G$-predictable if and only if  it admits a decomposition (see, e.g., Lemma 2.1 in \cite{pha10}),
\beqs
\xi_t(\cdot) &=& f_0(s, \cdot)\mathbf{1}_{\{t \leq \tau_{1}\}} + \sum_{n \geq 1} f_n(s, \cdot, \tau_1, I_{\tau_1}, \ldots, \tau_n, I_{\tau_n}) \mathbf{1}_{\{\tau_{n} < t \leq \tau_{n+1}\}}, \label{eqDecG}
\enqs
 where each $f_n$ is ${\cal P}^{W,B^{(1)}} \otimes {\cal B}(\R^{2n})$-measurable, ${\cal P}^{W,B^{(1)}}$ being the predictable $\sigma$-algebra corresponding to $({\cal W}_\cdot\vee {\cal B}^{(1)}_\cdot)$. Then, one proceeds as in \cite{phatan09}, by considering conditional controls and using a countable selection (one needs, in addition, a technical result similar to Lemma 3.2 in \cite{YZ} for the shifting procedure).}
\hfill$\square$
{\begin{Remark}
Our control problem is similar to the one in \cite{pha10} (see also \cite{bayzho}),
so that a similar approach  seems possible. However, it does not perfectly fit  that setting for several reasons. First, our controls $\alpha_k$ are measurable with respect to $\mathcal{G}_{\tau_k}$, whereas, in  \cite{pha10}, they are measurable with respect to $\mathcal{G}_{\tau_k^-}$. Second,   we have an infinite number of trading times $\tau_{k}$, whereas \cite{pha10} considers a finite number. Third,  we consider an infinite horizon, so the backward recursive approach cannot be employed.
\end{Remark}}
%
We  use  DPP to relate our original problem to a  continuous-time control problem.
For each $x\geq 0$, let $\Ac_{0}(x)$ be the set of couples of stochastic processes $(c_s,\pi_s)_{s\geq 0}$  such that
\begin{itemize}
\item[-]$(c_s)_{s\geq 0}$ is $(\mathcal{W}_s\vee\mathcal{B}^{(1)}_s)_{s\geq 0}$-predictable,  nonnegative, and has locally integrable trajectories;
\vspace{-.2cm}
\item[-]$(\pi_s)_{s\geq 0}$ is $(\mathcal{W}_s\vee\mathcal{B}^{(1)}_s)_{s\geq 0}$-predictable,  and has locally square-integrable trajectories;
\vspace{-.2cm}
\item[-]
 $x + \int_0^\cdot (-c_s ds +\pi_s (b_L ds + \sigma_L dW_s)) \geq 0.$
 \end{itemize}
By Lemma \ref{lemBeforetau1},  \reff{eqDPP0} becomes
\beq \label{eqDPP}
V(r) &=& \sup_{0\leq  a\leq r} \sup_{(c,\pi) \in \Ac_0(r-a)} \E\left[ \int_0^{\tau_1} e^{-\beta s} U(c_s) ds + e^{-\beta \tau_1} V\left(R_{\tau_1} \right) \right].
\enq
We rewrite the inner optimization problem in \eqref{eqDPP}, i.e.
\beq\label{innerproblem}
\sup_{(c,\pi) \in \Ac_0(r-a)} \E\left[ \int_0^{\tau_1} e^{-\beta s} U(c_s) ds + e^{-\beta \tau_1} V\left(R_{\tau_1} \right) \right].
\enq
Define (see Remark \ref{rem:general}(i) for explanations)
\beq\label{drift}
b_Y \; := \; \gamma^2 b_I + (1-\gamma^2) \frac{\rho b_L \sigma_I}{\sigma_L}, \ \ \ \ \
b_J \;:= \;(1-\gamma^2) \big( b_I - \frac{\rho b_L \sigma_I}{\sigma_L} \big),
\enq
and, given $x,y\geq 0$ and $(c,\pi)\in\mathcal{A}_0(x)$, define $J,\tilde{X}^{x,c,\pi},\tilde{Y}^{y}$ as solutions to
\begin{eqnarray}\label{XYt}
{dJ_t} &=& J_t \left (b_J dt + \sigma_I \sqrt{1-\rho^2} \sqrt{1-\gamma^2} dB^{(2)}_t\right), \ \ J_0=1,\\
d\tilde{X}_t &=& -{c}_t ds + {\pi}_t (b_L dt + \sigma_L dW_t), \ \ \ \tilde{X}_0=x,\\
\label{XY2t}
d\tilde{Y}_t &= &\tilde{Y}_t\left(b_Y dt+  \sigma_I \big(\rho dW_t+\sqrt{1-\rho^2} \ \gamma dB^{(1)}_t\big)\right), \ \ \ \tilde{Y}_0=y.
\end{eqnarray}
Then, for each $t\in[0,\tau_1)$, we have
$X_t = \tilde{X}^{r-\alpha_0, c, \pi}_t,  \ A_t= \tilde{Y}^{\alpha_0}_t\cdot J_t$.
Set
$
\mathcal{W}_\infty := \bigvee_{t\geq 0} \mathcal{W}_t,  \ \mathcal{B}^{(1)}_\infty := \bigvee_{t\geq 0} \mathcal{B}^{(1)}_t, \ \mathcal{B}^{(2)}_\infty := \bigvee_{t\geq 0} \mathcal{B}^{(2)}_t.
$
As $\tau_1$ is independent of $\mathcal{W}_\infty \vee\mathcal{B}^{(1)}_\infty\vee \mathcal{B}^{(2)}_\infty$ and has distribution $\Ec (\lambda)$, whereas $c$, $J$, $\tilde{X}^{x,c,\pi},\tilde{Y}^{y}$ are
$(\mathcal{W}_\infty \vee\mathcal{B}^{(1)}_\infty\vee \mathcal{B}^{(2)}_\infty)$-measurable, we have
\beqs
&&\E \left[ \int_{0}^{\tau_{1}} e^{- \beta s} U(c_s) ds + e^{- \beta \tau_{1}} V(R_{\tau_{1}})\ \Big|\ \mathcal{W}_\infty\vee\mathcal{B}_\infty^{(1)}\vee \mathcal{B}_\infty^{(2)}\right] \\
&=&\int_0^\infty \lambda e^{-\lambda t} \left(\int_{0}^{t} e^{- \beta s} U(c_s) ds + e^{- \beta t} V( \tilde{X}_t^{r-a,c,\pi}+J_t\cdot \tilde{Y}^{a}_t) \right) dt\\\
&=& \int_0^\infty e^{-\beta s} U(c_s) \int_s^\infty \lambda e^{- \lambda t} dt \,ds +   \int_0^\infty \lambda e^{-(\lambda+ \beta) t} V( \tilde{X}_t^{r-a,c,\pi}+J_t\cdot \tilde{Y}^{a}_t)   dt \\
&=&\int_0^\infty e^{- (\beta+\lambda) t} \left( U(c_t) + \lambda  V( \tilde{X}_t^{r-a,c,\pi}+J_t\cdot \tilde{Y}^{a}_t) \right)dt,
\enqs
where, in the second equality, we used Fubini's Theorem.
On the other hand, as  $J$ is independent of  $\mathcal{W}_\infty\vee \mathcal{B}^{(1)}_\infty$, whereas $c$, $\tilde{X}^{x,c,\pi},\tilde{Y}^{y}$ are
$(\mathcal{W}_\infty\vee \mathcal{B}^{(1)}_\infty)$-{measurable}, conditioning the equality above with respect to $\mathcal{W}_\infty\vee \mathcal{B}^{(1)}_\infty$, we get
\beqs
&&\E \left[ \int_{0}^{\tau_{1}} e^{- \beta s} U(c_s) ds + e^{- \beta \tau_{1}} V(R_{\tau_{1}})\ \Big|\ \mathcal{W}_\infty\vee\mathcal{B}_\infty^{(1)}\right] \\
&=&\E \left[ \int_0^\infty e^{- (\beta+\lambda) t} \left( U(c_t) + \lambda  V( \tilde{X}_t^{r-a,c,\pi}+J_t\cdot \tilde{Y}^{a}_t) \right)dt\ \Big|\ \mathcal{W}_\infty\vee\mathcal{B}_\infty^{(1)}\right],\\
&=&\int_0^\infty e^{- (\beta+\lambda) t} \left( U(c_t) + \lambda G_\gamma[V] (t,\tilde{X}^{r-a,c,\pi}_t,\tilde{Y}_t^a)\right)dt,
\enqs
where
$G_\gamma[V](t,x,y):=\E \left[ V(x+yJ_t)\right]$
 (the dependence on $\gamma$ coming from the dependence on $\gamma$ of  $J$).
Then, we can rewrite \eqref{innerproblem} as
\beq\label{innerproblem2}
\sup_{(c,\pi) \in \Ac_0(r-a)} \E\left[ \int_0^\infty e^{- (\beta+\lambda) t} \left( U(c_t) + \lambda G_\gamma[V] (t,\tilde{X}^{r-a,c,\pi}_t,\tilde{Y}_t^a)\right)dt \right].
\enq
It is useful to define $G_\gamma$ as a linear operator from the space $\mathcal{M}_1(\R_+;\R)$  of measurable functions with at most linear growth to the space of measurable functions  $\mathcal{M}(\R_+^3;\R)$\,:
\beq \label{defG}
G_\gamma: \ \mathcal{M}_1(\R_+;\R)&\longrightarrow &\mathcal{M}(\R_+^3;\R)\\
 \psi&\longmapsto &  G_\gamma[\psi](t,x,y):= \E\left[\psi(x+yJ_t) \right].\nonumber
\enq
Useful properties of $G_\gamma$ are listed in Proposition \ref{prop:G}.
\subsection{The auxiliary control problem} The optimization problem \eqref{innerproblem2} is a continuous, non autonomous, stochastic control problem over an infinite horizon that we call auxiliary problem.
One can apply the dynamic programming approach to this problem defining the same problem for generic initial data.
For each $t\geq 0$, consider the couples of stochastic processes $({c},{\pi})$ such that
\begin{itemize}
\item[(h1)$'$]$({c}_s)_{s\geq t}$ is $(\mathcal{W}_s\vee\mathcal{B}^{(1)}_s)_{s\geq t}$-predictable,  nonnegative, and has locally integrable trajectories,
\vspace{-.2cm}
\item[(h2)$'$]$({\pi}_s)_{s\geq t}$ is $(\mathcal{W}_s\vee\mathcal{B}^{(1)}_s)_{s\geq t}$-predictable, and has locally square-integrable trajectories,
 \end{itemize}
and define, for $x\geq 0$,
 $$\Ac_{t}(x):=\left\{(c,\pi)\ \mbox{fulfilling (h1)}'-\mbox{(h2)}'\  \big| \ x + \int_t^{\cdot} (-{c}_s ds +{\pi}_s (b_L ds + \sigma_L dW_s)) \geq 0\right\}.$$
 Let $(t,x,y)\in \R^3_+ $. Given $({c},{\pi}) \in\Ac_t(x)$, let  $(\tilde{X}_s^{t,x,{c},{\pi}})_{s\geq 0},(\tilde{Y}_s^{t,y})_{s\geq 0}$ be the solutions to
 \begin{eqnarray}\label{XYt0}
d\tilde{X}_s &=& -{c}_s ds + {\pi}_s (b_L ds + \sigma_L dW_s), \ \ \ \tilde{X}_t=x,\\
\label{XY2t0}
d\tilde{Y}_s &= &\tilde{Y}_s\left(b_Y ds+  \sigma_I \big(\rho dW_s+\sqrt{1-\rho^2} \ \gamma dB^{(1)}_s\big)\right), \ \ \ \tilde{Y}_t=y.
\end{eqnarray}
By definition of $\mathcal{A}_t(x)$, we have $\tilde{X}^{t,x,{c},{\pi}}\geq 0$. Moreover,    $\tilde{Y}^{t,y}\geq 0$.
Define the (auxiliary) value function
\beq\label{hatv}
\widehat{V}(t,x,y)\ :=\
\sup_{(c,\pi) \in \Ac_t(x)} \E\left[ \int_t^\infty e^{- (\beta+\lambda) (s-t)} \left( U(c_s) + \lambda G_\gamma[V] (s,\tilde{X}^{t,x,c,\pi}_s,\tilde{Y}_s^{t,y}\right)ds \right].
\label{defvhat}
\enq
Associating to every locally bounded function  $\hat{v}$ on $\R_+^3$,  the function, defined for $r\geq 0$,
$
[\Hc \hat{v}](r) := \sup_{0 \leq a \leq r} \hat{v}(0,r-a,a)$,
by   \eqref{eqDPP} and \eqref{defvhat}, we get
\beq\label{vhatv}
V (r)&=& [\Hc \widehat V](r), \ \ \ \forall r\geq 0.
\enq
The problems \eqref{hatv}-\eqref{vhatv} are coupled\,: $\widehat{V}$ is defined in terms of $V$ in \eqref{hatv} and $V$ is expressed in terms of $\widehat{V}$ in \eqref{vhatv}.
\begin{Remark}\label{rem:general}
\textbf{(i)}
The choices for the drifts $b_Y$ and $b_J$ in \eqref{drift} are motivated by the fact that we need
a couple of processes $(\tilde Y, J)$ such that: (i) $\tilde Y \cdot J=A$ on $[0,\tau_1)$, where $A$ is defined in \eqref{def:Z};  (ii)  $\tilde Y$ is $\G$-adapted; (iii) $J$ is independent of $\mathbb{G}$.
Therefore,  it is natural to consider the processes   \reff{XYt} and \reff{XY2t}, with $b_Y, b_J$ that can be chosen
under the constraint $b_J + b_Y$ $=$ $b_I$.
Define the constants
\begin{multline}\label{klyp}
k_{L,Y,p}\; := \;\sup_{u_L\in \R ,u_Y \in [0,1]} \Big\{\;p (u_L b_L + u_Y b_Y) \\- \frac{p(1-p)}{2} ( u_L^2 \sigma_L^2 + u_Y^2 \sigma_I^2(\rho^2 + \gamma^2(1-\rho^2)) + 2 \rho u_L u_Y \sigma_L \sigma_I)\Big\},
\end{multline}
\beq\label{kjp}
k_{J,p}& :=& \sup_{u_J \in [0,1]} \left\{p b_J u_J - \frac{p(1-p)}{2} \sigma_I^2 (1-\rho^2)(1-\gamma^2) u_J^2\;\right\}.
\enq
These constants naturally appear, respectively, in Lemma \ref{lemGrowthXY} and Proposition \ref{prop:G}-(v). {Combining these two results with \eqref{estimV}, one gets an estimate on the growth of $\widehat{V}$  (see \reff{Growthvhat} below) under the condition that $\beta>k_{L,Y,p}+k_{J,p}$.
In Lemma \ref{lemSplitting}, it is proved that, for our choice of $b_Y,b_J$,
\beq \label{hhj1}
 k_{L,Y,p}+k_{J,p} &=& k_p,
\enq
which is the minimum possible value of $ k_{L,Y,p}+k_{J,p}$.
So our drift choices enable us to treat the auxiliary problem without restrictions on $\beta$ other than Assumption \ref{ass:beta}.}

\smallskip
\textbf{(ii)}
The auxiliary problem \eqref{hatv}  is not autonomous, due to the dependence of $G_\gamma[V]$ on time (in general).
In the case of full observation ($\gamma=1$), one has $J \equiv 1$ and $G_1[V](t,x,y)=V(x+y)$, hence,
 consistent with \cite{APW}, we get an autonomous problem\,:\,\footnote{Our dynamic programming approach differs from \cite{APW}. Our approach seems to be the only one possible to deal with partial information ($\gamma<1$). Hence,  the differential problem we get is different from the one  in \cite{APW}; this difference remains  when our control problem coincides with one of \cite{APW} $(\gamma=1$ and power utility). Nevertheless, as  intuitively expected, our  problem is autonomous like the one in \cite{APW}, when the two control problems coincide.}
$$\sup_{(c,\pi) \in \Ac_0(x)} \E\left[ \int_0^\infty e^{- (\beta+\lambda) s} \left( U(c_s) +\lambda V\left(\tilde{X}_s^{0,x,{\pi},{c}}+\tilde{Y}_s^{0,y}\right) \right)ds\right].$$ Therefore, \emph{time inhomogeneity of our auxiliary problem is due to the lack of full information}.
We also note that we take,  in \eqref{hatv},  the discount $e^{-(\beta+\lambda)(s-t)}$ in place of the usual $e^{-(\beta+\lambda)s}$\: with this choice, we can  get rid of the exponential terms in the HJB equation.

\smallskip
\textbf{(iii)} The value function $\widehat V$ at time $t=0$ is the analogue of the value function of \cite{APW}, where the agent starts with \emph{given} initial liquid endowment $x$ and initial illiquid endowment $y$  (in \cite{APW}, $t=0$ is not a rebalancing time, so,  unlike  our case, the agent is not allowed to split the total endowment $r=x+y$ in a different proportion at $t=0$). 
In that regard, note that there is no loss of generality in assuming that $t=0$  is a trading time of the illiquid asset\,:
to treat the problem where $t=0$ is not a trading date for the illiquid asset and the initial endowment in liquid and illiquid are respectively $x$ and $y$, it suffices to not perform the first static optimization \eqref{vhatv}.

\end{Remark}

\subsection{HJB equation and viscosity characterization of $\widehat{V}$}\label{sec:visc}
This section  characterizes $\widehat{V}$ as unique constrained viscosity solution of an associated HJB equation. First, we state some qualitative properties of $\widehat{V}$. We omit the proof, which parallels \cite{FGtec}, where only the case $\gamma=0$ is considered.
\begin{Proposition}\label{prop:Vhat}
 For every $t\geq 0$, $\widehat{V}(t,\cdot, \cdot)$ is concave with respect to $(x,y)$ and  nondecreasing with respect to  $x$ and $y$. Moreover, it satisfies the boundary condition
 \beq \label{Bndryv}
\widehat{V}(t,0,y)&=& \E \left[\int_t^\infty e^{-(\beta+\lambda)(s-t)} \lambda G_\gamma[V](s,0,\tilde{Y}^{t,y}_s) ds\right], \ \ \ \forall t\geq 0, \ \forall y\geq 0.
\enq
Finally,  $\widehat V$ is continuous on $ \R_+^3$ and satisfies, for some $K_{\widehat{V}}>0$, the growth condition
\beq \label{Growthvhat}
0&\leq& \widehat V(t,x,y)\ \ \leq\ \ K_{\widehat V} e^{k_{J,p} t} (x+y)^p, \ \ \  \ \forall (t,x,y)\in \R_+^3.
\enq
\end{Proposition}
\medskip
\noindent
Let  $\Sc_2$ denote the space of real symmetric $2 \times 2$ matrices. Standard arguments of stochastic control (see, e.g., \cite[Ch.\,4]{YZ})  associate to $\widehat{V}$ the HJB equation\footnote{In the standard derivation of the HJB equation, one gets $-\lambda G_\gamma[V]$  in place of the third term. Therefore, the HJB equation is coupled with \eqref{vhatv}, as in \cite{phatan08, phatan09}. We  insert the nonlocal term \eqref{vhatv} directly in the equation.}
\beq\label{HJBv}
- \hat{v}_t  + (\beta+\lambda) \hat{v}  - \lambda G_\gamma[\Hc \hat{v}]- \sup_{c\geq 0, \pi \in \R} H_{cv}(y,D_{(x,y)} \hat{v},D^2_{(x,y)} \hat{v};\ c,\pi)&
 =& 0,
\enq
where, for $(y,q,Q)$ $\in$ $\R_+ \times \R^2 \times \Sc_2$, $c\geq 0, \pi \in \R$, the function $H_{cv}$ is defined as
\begin{multline*}
H_{cv}(y,q,Q;c,\pi)\\
 :=  \;U(c) + (\pi b_L - c) q_1 + b_Y y q_2+ \frac{\sigma_L^2 \pi^2}{2} Q_{11} +   \pi \rho \sigma_I \sigma_L y Q_{12}+ (\rho^2 + \gamma^2(1-\rho^2)) \frac{\sigma_I^2}{2} y^2 Q_{22}.
\end{multline*}
\begin{Remark}
 Equation \eqref{HJBv} has two nonlocal terms\,: $\mathcal{H}$ and $G_\gamma[\cdot]$. The first is due to illiquidity, the second  to partial observation and it disappears in the case of full observation (see Remark \ref{rem:general}(ii)).
\end{Remark}

\begin{Definition}\label{Def:visc} 
\rm
\ni \textbf{(1)}
An upper-semicontinuous (resp., lower-semicontinuous) function $v$ is  a \emph{viscosity subsolution} (resp., \emph{supersolution}) to \reff{HJBv} at $(t,x,y)\in \R_+^3$, if
\beqs
- \varphi_t(t,x,y)  + (\beta+\lambda) \varphi(t,x,y)  - \lambda G_\gamma[\Hc v](t,x,y) && \\- \sup_{c\geq 0, \pi \in \R}  H_{cv}(y,D_{(x,y)} \varphi(t,x,y),D^2_{(x,y)} \varphi(t,x,y);\ c,\pi) &\leq & 0 \;\;\;(\mbox{resp., } \geq 0),
\enqs
for each  $\varphi\in C^{1,2}(\R^3_+;\R)$ such that $\varphi(t,x,y) = v(t,x,y)$ and $\varphi$ $\geq$ $v$ (resp., $\leq$) on $[t,t+\eps) \times \mathcal{O}$, for some neighborhood $\mathcal{O}\subset \R_+^2$ of $(x,y)$ and some $\varepsilon > 0$.
\smallskip

 \textbf{(2)}
A continuous function $v$ is  a \emph{constrained viscosity solution} to \reff{HJBv}, if it is a subsolution at all $(t,x,y)\in \R_+^3$, a supersolution at all  $(t,x,y)\in \R_+\times (0,+\infty) \times \R_+$, and satisfies the boundary condition
\beq\label{bdrcond}
\hat {v}(t,0,y)&=& \E \Big[\int_t^\infty e^{-(\beta+\lambda)(s-t)} \lambda G_\gamma[\mathcal{H}\hat v](s,0,\tilde{Y}^{t,y}_s) ds\Big], \ \ \ \forall t\geq 0, \ \forall y\geq 0.
\enq
\end{Definition}

\begin{Remark}\label{ParabolicVisc}
\textbf{(i)} To simplify proofs, we  use a definition of viscosity solution that differs from the usual one. In the standard definition (see \cite{CIL92}), the test functions $\varphi$ stay above $v$ in a neighborhood of $(t,x,y)$, whereas the property in our case is required only for $s\geq t$. Our definition is more restrictive, as we enlarge the set of test functions. Nevertheless, the two definitions are equivalent  if the  comparison principle for viscosity solutions holds for  the standard definition (see \cite{Juu01}).

{\textbf{(ii)}
The notion of \emph{constrained} viscosity solution is specific to  our stochastic control problem. The boundaries $\mathcal{B}_1:=\{(x,y)\in\R^2_+\, :\, x=0, \ y> 0\}$ and $\mathcal{B}_2:=\{(x,y)\in\R^2_+\, :\,  x> 0, \ y=0\}$ are both absorbing (that is, state trajectories staring on $\mathcal{B}_i$  remain on $\mathcal{B}_i$), although differently. When the initial state is on  $\mathcal{B}_2$, the control problem degenerates to a one dimensional control problem whose HJB equation is \eqref{HJBv} restricted to $\mathcal{B}_2$. For this reason, the value function must satisfy  viscosity sub- and supersolution properties. When the initial state is on $\mathcal{B}_1$, the  control problem vanishes, as $\mathcal{A}_t(0)=\{(0,0)\}$, and the natural condition to impose is of Dirichlet type. However, the  viscosity subsolution property  holds on $\mathcal{B}_1$\,: for this reason, we require it in our definition, even if it is  redundant for the purpose of establishing a comparison property.}
\end{Remark}

%
%
%

\begin{Theorem} \label{thmVisc}
$\widehat{V}$ is the unique constrained viscosity solution to  \eqref{HJBv}
satisfying  \reff{Growthvhat}.
\end{Theorem}
{\bf Proof.} The proof that $\widehat{V}$ is a viscosity subsolution on  $\R_+^3$ and a viscosity supersolution on $\R_+\times (0,+\infty)^2$ parallels, e.g.,  \cite[Ch.\,4,\, Th.\,5.2]{YZ}
{(see also \cite{BS13} for an approach via the stochastic Perron method, which requires only a viscosity comparison  property and no Dynamic Programming Principle)}. The Dirichlet boundary condition \eqref{bdrcond} follows from  \eqref{Bndryv} and \eqref{vhatv}. The growth condition \reff{Growthvhat} was  proven.
When  $y=0$, as noted in Remark \ref{ParabolicVisc}(ii), the control problem is  one-dimensional  and $\widehat{V}$ is a viscosity supersolution by standard arguments.

Uniqueness is a consequence of the comparison principle (Proposition \ref{Prop:comparison}), whose proof parallels  \cite{FGtec}, where only the case $\gamma=0$ is treated. \hfill$\square$

\begin{Proposition}\label{Prop:comparison}
Let $v_1$ (resp., $v_2$) be a viscosity subsolution (resp., supersolution) to \reff{HJBv} on $\R_+ \times (0,+\infty) \times \R_+$. Assume that $v_1$, $v_2$ satisfy the growth condition \reff{Growthvhat} and
\beq \label{eqBndrySub}
v_1(t,0,y) &\leq& \E \Big[\int_t^\infty e^{-(\beta+ \lambda)(s-t)} \lambda G_\gamma[\Hc v_1](s,0,Y^{t,y}_s) ds\Big]
\enq
(resp., $\geq$ for $v_2$). Then $v_1$ $\leq$ $v_2$ on $\R_+^3$.
\end{Proposition}

{

\subsection{{Optimal policy in the illiquid asset}}\label{sec:optpol}

The results  obtained above characterize the optimal allocation policy in the illiquid asset.
If $\widehat V$ can be computed numerically as a viscosity solution of \eqref{HJBv} (see Section \ref{Sec:num}),  then the optimal allocation policy $(\alpha_k^*)_{k\in\N}$ in the illiquid asset can be derived.
 At time $t=0$,  \eqref{vhatv} implies that  $\alpha_0^*$ is an optimal allocation in the illiquid asset if and only if
$
\alpha_0^*\in \mbox{argmax}_{\,0\leq a\leq r}\ \widehat{V}(0,r-a, a).
$
Note  that the case $\alpha_0^*=0$ cannot be ruled out   at this stage.
The property can be generalized to the random trading dates $\tau_k$\,:   {$\alpha_{k}^*$ is an  optimal allocation in the illiquid asset if and only if }
\beq\label{alphaopt}
\alpha_k^*&\in& \mbox{argmax}_{\,0\leq a\leq {R}_{\tau_k}}\widehat{V}(0,R_{\tau_k}-a, a), \ \ \ \ k\in\N,
\enq
as a consequence of   the Markov property. We omit the proof for brevity.

\section{Power utility}\label{Sec:power}
\setcounter{equation}{0}
{To characterize  optimal policies, smoothness of  $\widehat{V}$ is needed.
Unfortunately, the HJB equation \eqref{HJBv} is degenerate, as  the control problem has two state variables and a one dimensional Brownian motion, and   the  regularity theory for PDEs does not cover that case. Hence, to proceed, we assume that  the utility function $U$ is of power type, so that, as in \cite{APW, ST04},  the problem reduces to one spatial dimension.}
\begin{Assumption}\label{ass:power}
$U(c) \;= \;\frac{c^p}{p}, \ p\in(0,1).$
\end{Assumption}
Assumption \ref{ass:power}  holds from here on.

 \begin{Remark}
 When $p\leq 0$, the  problem is investigated in \cite{APW} assuming  full observation of the illiquid asset, which corresponds  to our case $\gamma=1$.
 \end{Remark}

\subsection{Reduction to one spatial variable} \label{subsec:pow1}

\begin{Proposition}\label{prop:power}
There exists $K_V>0$ such that
\beq \label{Phi_0}
V(r) \;=\; {K_V} r^p,  \ \ \ \forall r\geq  0.
\enq
Hence
\begin{equation}\label{eqGgrowthG2}
G_\gamma[V](t,\xi x,\xi y)\ =\  \xi ^p G_\gamma[V](t,x, y), \ \ \ \forall t\geq 0, \ \forall (x,y)\in \R_+^2,  \ \forall  \xi\geq 0,
\end{equation}
 and
\beq\label{VVV}
\widehat{V}(t,\xi x,\xi y) \;=\;\xi^p\ \widehat V(t,x,y), \ \ \ \forall \xi\geq 0, \ \forall x,y\geq 0.
\enq
\end{Proposition}
\textbf{Proof.}
By linearity of the state equations,
$\mathcal{A}(\xi r)= \xi \mathcal{A}(r)$ for every
$\xi\geq 0 $ and  $r\geq 0.$
By homogeneity of $U$, we  get \eqref{Phi_0}; then \eqref{eqGgrowthG2} follows from \eqref{eqGgrowthG1}.
Again by linearity,
$\mathcal{A}_t(\xi x)=\xi \mathcal{A}_t(x)$ for every $ \xi\geq 0$ and  $x\geq0$.
Then \eqref{VVV} follows  from \eqref{Phi_0} and \eqref{eqGgrowthG2}.\hfill$\square$\\

Under Assumption \ref{ass:power}, \eqref{VVV} becomes
\beq
\widehat{V}(t,x,y)\ =\
\sup_{(c,\pi) \in \Ac_t(x)} \E\left[ \int_t^\infty e^{- (\beta+\lambda) (s-t)} \left( \frac{c_s^p}{p} + \lambda G_\gamma[V] (s,\tilde{X}^{t,x,c,\pi}_s,\tilde{Y}_s^{t,y}\right)ds \right].
\label{defvhat1}
\enq
If  $y=0$, then $\tilde{Y}_s^{t,y}\equiv 0$. So,
 by Proposition \ref{prop:power},
 \beq\label{degcon}
 \widehat V (t,x,0) &=&\!\!\! \sup_{({c},{\pi}) \in \Ac_t(x)} \E \left[ \int_t^\infty e^{- (\beta+\lambda) (s-t)} \left( \frac{c_s^p}{p} + \lambda {K_V} [\tilde{X}_s^{t,x,c,\pi}]^p \right) ds\right].
 \enq
This is a standard homogeneous  Merton type problem, for which $\widehat{V}(t,x,0)= K_0 x^p$ for some $K_0>0$. We omit to treat this case.\footnote{
A necessary and sufficient condition excluding the case $y=0$ is given in Proposition \ref{propCNS} below.}

If $y>0$, then $\tilde{Y}_s^{t,y}>0$ for every $s\geq t$. In view of \eqref{VVV}, we consider a new state process $Z$, defined as the ratio of $\tilde{X}$ and $\tilde{Y}$. More precisely, let $(c, \pi)\in\mathcal{A}_t(x)$, $\tilde{X}:=\tilde{X}^{t,x,c,\pi}$, $\tilde{Y}:=\tilde{Y}^{t,y}$, $Z_s := \frac{\tilde{X}_s}{\tilde Y_s}$. It\^o's formula yields
\beq \label{eqDynZ1}
dZ_s &=& -{\hat{c}}_s ds+ {\hat{\pi}}_s (\widehat K_1 ds+ \sigma_L dW_s) + Z_s ( \widehat K_2 ds + K_\gamma dB^{(1)}_s ),
\enq
where
\beq
\begin{cases}
{\hat{c}}_s := \frac{c_s}{\tilde{Y}_s},   \\ \hat{\pi}_s := \frac{\pi_s}{\tilde{Y}_s} - \frac{\rho \sigma_I}{\sigma_L} \frac{\tilde{X}_s}{\tilde{Y}_s},\\
\widehat K_1 := b_L-\rho\sigma_I\sigma_L,\\
\widehat K_2 := \gamma^2 \left(-b_I + \frac{\rho b_L \sigma_I}{\sigma_L}+(1-\rho^2)\sigma_I^2\right),\\
K_\gamma :=  - \sigma_I \gamma \sqrt{1-\rho^2}.
\end{cases}
\enq
Then,  by Proposition \ref{prop:power}, and for
$f_\gamma(t,z) := G_\gamma[\xi\mapsto \xi^p](t,z,1),$
 \eqref{defvhat1} becomes
\beq
\widehat{V}(t,x,y)\ =\
\sup_{(c,\pi) \in \Ac_t(x)} \E\left[ \int_t^\infty e^{- (\beta+\lambda) (s-t)} (\tilde{Y}_s)^p \left(\frac{{\hat{c}_s}^p}{p} + \lambda {K_V} f_\gamma (s,Z_s)\right)ds \right] .
\label{eqControlZ1}
\enq
In order to simplify \eqref{eqControlZ1}, we change  probability measure.
Let
 $\widehat \P$ be the measure  with density process $\frac{(\tilde{Y}_s)^p}{\E [(\tilde{Y}_s)^p]}$.
Under  $\widehat \P$, the processes
$\widehat W_s := W_s - p \rho \sigma_I s$\, and\,   $\widehat B^{(1)}_s := B^{(1)}_s - \gamma \sqrt{1-\rho^2} \ s$ \, are  Brownian motions and the dynamics of $Z$ can be written as
\beq\label{ZZ0}
dZ_s &=& -{\hat{c}}_s ds+ {\hat{\pi}}_s (K_1 ds+ \sigma_L d \widehat W_s) + Z_s ( K_2 ds + K_\gamma d \widehat B^{(1)}_s ),
\enq
where
\beq
K_1 \;:=\;b_L-\rho\sigma_I\sigma_L(1-p),\ \ \ \
K_2\;:=\; \displaystyle{\gamma^2 \left(-b_I + \frac{\rho b_L \sigma_I}{\sigma_L}+(1-\rho^2)(1-p)\sigma_I^2\right)}.
\enq
Moreover, \eqref{eqControlZ1} becomes
\beq  \label{eqControlZ2}
 \widehat{V}(t,x,y)\ =\  y^p\cdot
\sup_{(c,\pi) \in \Ac_t(x)}  \ \widehat{\mathbb{E}} \left[ \int_t^\infty e^{- K_{\lambda} (s-t)}\left(\frac{\hat{c}_s^{p}}{p}+ \lambda {{K_V}}f^{\gamma}(s,Z_s )\right)ds\right],
\enq
where $\widehat{\mathbb{E}}$ is the expectation under  $\widehat{\mathbb{P}}$ and
$$K_\lambda\;:=\; \displaystyle{\beta+\lambda +\frac{(\rho^2+\gamma^2(1-\rho^2))\sigma_I^2}{2}p(1-p)-p\rho \frac{b_L\sigma_I}{\sigma_L}-\gamma^2 p\left(b_I-\frac{\rho b_L\sigma_I}{\sigma_L}\right)}.
$$
Given $z\geq 0$, we consider \eqref{ZZ0} with initial datum $z$ as a controlled equation  with controls $(\hat{c},{\hat{\pi}})$, where\footnote{The filtration generated by $(\widehat W, \widehat {B}^{(1)})$ is the same as the filtration generated by  $(W, {B}^{(1)})$.}
\begin{itemize}
\item[(h1)$''$]$({\hat{c}_s})_{s\geq t}$ is $(\mathcal{W}_s\vee\mathcal{B}^{(1)}_s)_{s\geq t}$-predictable, nonnegative, and has locally integrable trajectories;
\item[(h2)$''$]$({\hat{\pi}}_s)_{s\geq t}$ is $(\mathcal{W}_s\vee\mathcal{B}^{(1)}_s)_{s\geq t}$-predictable, and has locally square-integrable trajectories.
 \end{itemize}
 Let $Z^{t,z,\hat{c},{\hat{\pi}}}$ be the solution to \eqref{ZZ0}, starting from $z$ at time $t$ and under a control $(\hat{c},{\hat{\pi}})$ fulfilling $\mbox{(h1)}''-\mbox{(h2)}''$. Set
$$\hat{\mathcal{A}}_t(z):= \{(\hat{c},{\hat{\pi}}) \ \mbox{fulfilling (h1)}''-\mbox{(h2)}'' \ | \ Z^{t,z,\hat{c},{\hat{\pi}}}\geq 0\},$$
and consider the one-dimensional stochastic control problem
\beq\label{def_phi}
\Phi(t,z):=\sup_{(\hat{c},{\hat{\pi}}) \in \hat{\mathcal{A}}_t(z)}\ \widehat{\mathbb{E}} \left[ \int_t^\infty e^{- K_{\lambda} (s-t)}\left(\frac{{\hat{c}_s}^p}{p}+ \lambda {{K_V}}f^{\gamma}(s,Z^{t,z,\hat{c},{\hat{\pi}}}_s )\right)ds\right].
\enq
\begin{Proposition}\label{prop:conn}
Let $x\geq 0,$  $y>0$ and $z:=x/y$.
\begin{enumerate}
\item $
(c,\pi)\in \mathcal{A}_t(x)
$ if and only if $(\hat{c},\hat{\pi})\in \hat{\mathcal{A}}_t(z)$, where
$
{\hat{c}}_s = \frac{c_s}{\tilde{Y}^{t,y}_s},  \  \hat{\pi}_s = \frac{\pi_s}{\tilde{Y}^{t,y}_s} - \frac{\rho \sigma_I}{\sigma_L} \frac{\tilde{X}^{t,x,c,\pi}_s}{\tilde{Y}^{t,y}_s}.
$
\item
 $
(c^*,\pi^*)\in \mathcal{A}_t(x)
$ is optimal for \eqref{defvhat1} if and only if $(\hat{c}^*,\hat{\pi}^*)\in \hat{\mathcal{A}}_t(z)$ is optimal for \eqref{def_phi}, where
$
{\hat{c}^*}_s = \frac{c^*_s}{\tilde{Y}^{t,y}_s}  \    \hat{\pi}^*_s = \frac{\pi^*_s}{\tilde{Y}^{t,y}_s} - \frac{\rho \sigma_I}{\sigma_L} \frac{\tilde{X}^{t,x,c^*,\pi^*}_s}{\tilde{Y}^{t,y}_s}.
$

\item
$\widehat{V}(t,x,y) = y^p \ \Phi(t, z)$, for every $t\geq 0$.
\end{enumerate}
\end{Proposition}
\textbf{Proof.} {All the claims follow from the arguments above.}\hfill$\square$

\medskip

In view of Proposition \ref{prop:conn}, from here on we study the optimization problem \eqref{def_phi}.
Denote by $\mathcal{M}_p(\mathbb{R}^2_+,\mathbb{R})$ the space of measurable functions $\psi$ such that $|\psi(t,z)|\leq C_0(1+|z|)^p$, and consider the nonlinear functional
$
\mathcal{H}_0:  \mathcal{M}_p(\mathbb{R}^2_+,\mathbb{R})  \rightarrow  \mathbb{R},$
$\psi\mapsto \mathcal{H}_0[\psi]:=\sup_{z\geq 0} \ \frac{\psi(0,z)}{(1+z)^p}.$
Then $
V(r)=\mathcal{H}[\widehat{V}](r)=\mathcal{H}_0[\Phi] r^p$,
so that
\beq\label{phi0}
{K_V}=\mathcal{H}_0[\Phi],
\enq
where $K_V$ is the constant in \eqref{Phi_0}.
As $\hat{\mathcal{A}}_t(0)=\{(c,\pi)\equiv (0,0)\}$,
we get the boundary condition for $\Phi$
\beq\label{bbb}
\Phi(t,0)\ = \ {K_V}\int_t^\infty e^{-K_\lambda(s-t)}\lambda f_\gamma(s,0) ds.
\enq
By \eqref{Growthvhat} and Proposition \eqref{prop:conn}(3),\footnote{It could also be proven by dealing directly with the control problem \eqref{def_phi}.} we get the  growth condition for $\Phi$
\beq\label{newgrowth}
\Phi(t,z) &\leq&  K_{\widehat V} e^{k_{J,p} t}(1+z)^p.
\enq
The HJB equation associated to \eqref{def_phi} is
\beq\label{HJBphi*}
- \varphi_t + K_\lambda \varphi -  \lambda {K_V} f_\gamma(t,z) -\sup_{\hat{c}\geq 0,\,\hat{\pi}\in \R}{H}^0_{cv}(z,\varphi,\varphi_z,\varphi_{zz})&=&0,
\enq
where
\begin{multline*}
{H}^0_{cv}(z,\varphi,\varphi_z,\varphi_{zz};\hat{c},\hat{\pi})= \frac{\hat{c}^p}{p}-\hat{c}\varphi_z+ K_1 \hat{\pi}\varphi_z+\frac{1}{2} \sigma_L^2\hat{\pi}^2\varphi_{zz}
+K_2 z\varphi_z+\frac{1}{2}K_\gamma^2z^2\varphi_{zz}.
\end{multline*}
By \eqref{phi0}, we can replace ${K_V}$ with $\mathcal{H}_0[\varphi]$ in \eqref{HJBphi*} and also consider  the equation with a nonlocal term
 \beq\label{HJBphi**}
- \varphi_t +K_\lambda \varphi  - K_2 z \varphi_z - \lambda f_\gamma(t,z) \mathcal{H}_0[\varphi]-\sup_{\hat{c}\geq 0,\,{\hat{\pi}}\in \R}\!\!\!{H}^0_{cv}(z,\varphi,\varphi_z,\varphi_{zz})\ =\ 0.
\enq
Similarly,  we can replace  ${K_V}$ with $\mathcal{H}_0[\varphi]$ in \eqref{bbb} and get an implicit nonlocal  boundary condition\,:
\beq\label{bbb1}
\Phi(t,0)\  =\ \mathcal{H}_0[\Phi]\ \int_t^\infty e^{-K_\lambda(s-t)}\lambda f_\gamma(s,0) ds.
\enq

\begin{Proposition}\label{prop:visc2} The function
 $\Phi$  is the unique {continuous} viscosity solution\ \footnote{The definition of viscosity solution is analogous to Definition \ref{Def:visc}(1).} over $\mathbb{R}_+\times (0,+\infty)$ to \eqref{HJBphi*}
  fulfilling  \eqref{bbb} and \eqref{newgrowth}.
Equivalently,  $\Phi$  is the unique {continuous} viscosity solution over $\mathbb{R}_+\times (0,+\infty)$ to \eqref{HJBphi**}
  fulfilling  \eqref{bbb1} and  \eqref{newgrowth}.

\end{Proposition}

\textbf{Proof.}
The first  fact
follows from Proposition \ref{prop:conn}(3)  and Theorem \ref{thmVisc}.
The  equivalence between the equations follows from uniqueness and \eqref{phi0}.
\hfill$\square$

\subsection{Smoothness of the value function}
This subsection  shows that the value function $\Phi$ is smooth.
As the classification of the HJB equation \eqref{HJBphi*} is sensitive to $\gamma$, we need to distinguish, from the point of view of PDE theory, the cases  $\gamma=0$ and $\gamma\neq 0$.
Due to the presence of the term $\frac{K_\gamma^2}{2} z^2 \varphi_{zz}$,
in the case $\gamma\neq 0$ ($K_\gamma>0$), the PDE is a fully nonlinear {(locally)} {nondegenerate} parabolic equation, whereas, in the case $\gamma=0$ ($K_\gamma=0$), it is {degenerate}.
In both cases, we prove that the solution is sufficiently  smooth  to construct optimal policies in feedback form. The degenerate case, $\gamma=0$, is investigated by means of the dual problem  illustrated in Remark \ref{Rem:dual} and discussed in detail in \cite[Sec.\,6]{FGGdual}. {For the nondegenerate  case, $\gamma\neq 0$, we first localize the equation by restricting the set of controls to a compact one  and then apply a result by Krylov (see  Appendix).}
Hereafter,  we  denote by $C^{1,k}$ the class of functions which are once differentiable with respect to the time variable and $k$-times differentiable with respect to  spatial variable,  with  continuous derivatives.

\smallskip

\begin{Theorem} \label{prop:reg}
$\Phi\in C^{1,3}(\mathbb{R}_+\times (0,+\infty);\R)$ and $\Phi_z\in C^{1,2}(\mathbb{R}_+\times (0,+\infty);\R)$, with
 $\Phi_z>0$ and  $\Phi_{zz}<0$ over $\mathbb{R}_+\times (0,+\infty)$.
\end{Theorem}
\textbf{Proof.}
We prove the claim in the case $\gamma\neq 0$, referring to  \cite{FGGdual} and Remark \ref{Rem:dual} below for the case $\gamma=0$.
Given $(\bar{t},\bar{z})\in\R_+\times (0,+\infty)$ and $\varepsilon\in(0,\bar{z})$,  consider  $D^\varepsilon(\bar{t},\bar{z})$, defined in \eqref{D_epsilon}, and let ${\cal P}(D_\varepsilon (\bar{t},\bar{z}))$ be the parabolic boundary of $D_\varepsilon (\bar{t},\bar{z})$\,:
$$
{\cal P}(D_\varepsilon (\bar{t},\bar{z}))\ :=\ \{\bar{t}+\varepsilon\}\times [\bar{z}-\varepsilon,\bar{z}+\varepsilon]
\cup  [\bar{t},\bar{t}+\varepsilon]\times  \{\bar{z}-\varepsilon,\bar{z}+\varepsilon \}.
$$
{By Propositions \ref{prop:visc2}} and \ref{prop:HJBv}, and by standard comparison results for viscosity solutions (see e.g. {\cite{CIL92,FS06}}), ${\Phi}$ is the unique continuous viscosity solution on $D_\varepsilon (\bar{t},\bar{z})$ to the  HJB equation
\eqref{HJBphi3} - the same as \eqref{HJBphi*}, but with constraints on the set of the variables $\hat{c},{\hat{\pi}}$ -  with Dirichlet continuous boundary condition
\begin{equation}\label{eq:BCforHJBdualforreg}
\varphi\ =\ {\Phi}, \qquad \mbox{on} \quad {\cal P}(D_\varepsilon (\bar{t},\bar{z})).
\end{equation}
On the other hand, by Theorem \ref{teo:krylov}, there exists  a solution $C^{1,2}(D_\varepsilon (\bar{t},\bar{z});\R)$  to \eqref{HJBphi3} with boundary condition \eqref{eq:BCforHJBdualforreg}. As such a solution needs to  be  a viscosity solution,
\beq\label{Phic2}
\Phi \ \in\  C^{1,2}(D_\varepsilon (\bar{t},\bar{z});\R).
\enq
Moreover, by Lemma \ref{lemma:sub}, we  have for  $(t,z)\in D_\varepsilon (\bar{t},\bar{z})$
\beq\label{ppw}
\Phi_z(t,z)\  \geq  \ m_\varepsilon \ >\ 0, \ \ \ \ \ \ \Phi_{zz}(t,z)\  \leq  \ -\delta_\varepsilon\ <\ 0.
\enq
Due to \eqref{Phic2}-\eqref{ppw} and Proposition \ref{prop:visc2}, and by arbitrariness of $(\bar{t},\bar{z})\in\R_+\times(0,+\infty)$,  the function $\Phi$ is a classical solution  to \eqref{HJBphi*} and $\Phi_z>0$, $\Phi_{zz}<0$ in $\mathbb{R}_+\times (0,+\infty)$. As a consequence
 the supremum in \eqref{HJBphi*} can be made explicit, so that $\Phi$ satisfies, in the classical sense,
\beq\label{HJBphi5}
- \Phi_t + K_\lambda\Phi- K_2 z \Phi_z - \lambda {K_V} f_\gamma(t,z)-\widetilde{U}(\Phi_z)+\frac{1}{2}\frac{K_1^2}{\sigma_L^2}
\frac{\Phi_z^2}{\Phi_{zz}}- \frac{K_\gamma^2}{2} z^2 \Phi_{zz}&=&0,
\enq
where  $\widetilde U(w):=\frac{1-p}{p}\ w^{-\frac{p}{1-p}},$  $w>0$, is the Legendre transform  of $U$. By
Lemma \ref{lemDiffVisc},  we  differentiate \reff{HJBphi5} and deduce that $\Phi_z$ is a viscosity solution to
\beq \label{pas2}
-g_t + \Big(K_\lambda + K_2 + \frac{K_1^2}{\sigma_L^2}\Big) g + (K_2-K_\gamma^2) z g_z + \tilde{U}'(g) g_z && \nonumber\\
- \Big(\frac{K_\gamma^2}{2} z^2 + \frac{K_1^2}{2 \sigma_L^2} \frac{g^2}{g_z^2}\Big) g_{zz} + \lambda {K_V} (f_\gamma)_z &=& 0,
\enq
with Dirichlet continuous boundary condition
$
g = {\Phi_z}$ on  $ {\cal P}(D_\varepsilon (\bar{t},\bar{z})).
$
Again, by standard comparison results for viscosity solutions, the function $\Phi_z$  is the unique viscosity solution to this problem. On the other hand,  {Theorem \ref{teo:krylov}\footnote{
Note that  $\tilde{U}'(g) g_z$ and ${g^2}/{g_z^2}$ are not well-defined for $g\leq 0$ or $g_z=0$. We can use \reff{ppw} to replace these terms by  functions of $(g,g_z)$ everywhere defined, with as much smoothness as needed, coinciding with $(g,g_z)$ whenever $m_\eps \leq g \leq M_\eps$, $g_z \leq - \delta_\eps$, and satisfying the assumptions of Theorem \ref{teo:krylov}.}}} implies  that this problem admits a  $C^{1,2}(D_\varepsilon (\bar{t},\bar{z});\R)$ solution. As before, we deduce  $\Phi_{z}\in C^{1,2}(D_\varepsilon (\bar{t},\bar{z});\R)$,  hence $\Phi\in C^{1,3}(D_\varepsilon (\bar{t},\bar{z});\R)$. We finally get the claim by arbitrariness  of $(\bar{t},\bar{z})\in\R_+\times (0,+\infty)$. \hfill\ep
\begin{Remark}\label{Rem:dual}
The equation \eqref{HJBphi*}, when $\gamma=0$, is degenerate, as the second order term can vanish, but $\Phi$ is still a viscosity solution of it. Motivated by the fact that  $\Phi_z>0$ and $\Phi_{zz}<0$, one rewrites the equation as
\beq\label{HJBphi10}
- \varphi_t + K_\lambda\varphi- K_2 z \varphi_z - \lambda {K_V} f_0(t,z)-\widetilde{U}(\varphi_z)+\frac{1}{2}\frac{K_1^2}{\sigma_L^2}
\frac{\varphi_z^2}{\varphi_{zz}}&=&0.
\enq
Define the dual function $\psi(t,w):=\sup_{z\geq 0} \{\varphi(t,z)-wz\}$, $w>0$. From  \eqref{HJBphi10}, one formally gets the equation for $\psi$
\begin{equation}\label{eq:HJBdualpsi}
-\psi_t+{K_\lambda}\psi+(K_2-K_\lambda)w\psi_w-\frac{K^2_1}{2\sigma_L^2}w^2\psi_{ww}-\widetilde{U}(w)-\lambda K_V f_0(t,-\psi_w)=0
\end{equation}
(boundary and growth conditions  for $\varphi$ also have a dual counterpart for $\psi$). The equation \eqref{eq:HJBdualpsi} is semilinear and (locally) nondegenerate, so the PDE theory for classical solutions can be used. Moreover, \eqref{eq:HJBdualpsi}   is still  of HJB type, itself associated with another (dual) control problem. Once  the existence of a sufficiently smooth   solution $\Psi$  to \eqref{eq:HJBdualpsi} is derived, one may try to show that $\widetilde{\Psi} (t,z):=\inf_{w>0}\{\Psi(t,w)+zw\}$ is  a classical solution to \eqref{HJBphi10}, and identify it with $\Phi$. All these steps are nontrivial (see \cite{FGGdual}).
\end{Remark}

\subsection{Closed loop equation}
The candidate optimal feedback maps provided  by  maximization
in the HJB equation \eqref{HJBphi*}, for $z>0$, and by the fact that $\hat{\mathcal{A}}_t(0)=\{(0,0)\}$, for $z=0$, are

\beq\label{def:feed}
\widehat{C}^*(s,z)\;=\;\begin{cases}({U}')^{-1}\left(\Phi_z(s,z)\right), \ \ \ \mbox{if} \ z>0,\\
0, \ \ \ \ \ \ \ \ \ \ \ \  \ \ \ \ \ \ \ \ \ \ \ \ \mbox{if} \ z=0,
\end{cases} \ \
\widehat{\Pi}^*(s,z)\;=\;\begin{cases}-\frac{K_1\Phi_z(s,z)}{\sigma_L\Phi_{zz}(s,z)}, \ \ \ \mbox{if} \ z>0,\\
0, \ \ \ \ \ \ \ \ \ \ \ \, \ \ \ \ \    \mbox{if} \ z=0.
\end{cases}
\enq
These maps are measurable. Moreover,
due to Theorem \ref{prop:reg},  $\widehat{C}^*(s,\cdot)$ and  $\widehat{\Pi}^*(s,\cdot)$ are locally Lipschitz continuous  in $(0,+\infty)$, uniformly in  $s\in[0,T]$, for all $T>0$.
The associated closed loop equation is
\beq\label{CLE}
\begin{cases}
dZ_s\ =\  -\widehat{C}^*(s,Z_s)ds+ \widehat{\Pi}^*(s,Z_s) \left( K_1 ds+ \sigma_L d\widehat W_s \right) + Z_s (  K_2 ds + K_\gamma d\widehat B^{(1)}_s ),\\
Z_t\ =\ z.
\end{cases}
\enq

\begin{Proposition} \label{prop:CLE-Z}
For every $(t,z)\in\R_+^2$, there exists a unique nonnegative strong solution $Z^{t,z,*}$ to \eqref{CLE} in   $(\Omega,\mathcal{F},\widehat{\mathbb{P}})$.
\end{Proposition}

\ni{\bf Proof.}
\emph{Existence.} If $z=0$, the claim follows by setting  $Z^{t,z,*}\equiv 0$. Let $z>0$ and $T>0$. The local Lipschitz continuity of $\widehat{C}^*(s,\cdot), \widehat{\Pi}^*(s,\cdot)$ and standard SDEs theory (see, e.g., \cite[Ch.\,5,\,Th.\,2.9]{KS}) give, for each $\varepsilon\in(0,z)$,  the existence of a unique solution $Z^{t,z,\varepsilon}\in [\varepsilon,\varepsilon^{-1}]$ in the stochastic  interval $[t,\tau^T_\varepsilon)$, where $\tau^T_\varepsilon$ is implicitly defined as
$
\tau^T_\varepsilon:=\inf\ \{s\in[t,T]\ | \ Z_s^{t,z,\varepsilon}\leq \varepsilon \mbox{ or } Z_s^{t,z,\varepsilon}\geq \varepsilon^{-1}\},
$
with the convention $\inf\emptyset=T$.
If $\varepsilon<\varepsilon'$,  we have $\tau^T_\varepsilon>\tau^T_{\varepsilon'}$ and
\beq\label{ppp}
 Z_s^{t,z,\varepsilon}&\equiv& Z_s^{t,z,\varepsilon'} \ \ \mbox{on}\ \  [t,\tau^T_{\varepsilon'}), \ \ \  \ \ \forall \ 0<\varepsilon<\varepsilon'.
\enq
Set $\tau^T:=\lim_{\varepsilon\downarrow 0}\tau^T_\varepsilon.$
By \eqref{ppp}, there exists a unique solution  $Z^{t,z,*}\geq 0$ in the interval $[t,\tau^T)$. We  show that $Z^{t,z,*}$  can be extended to  $[t,T]$, which implies the claim, by arbitrariness of $T$.
By a Girsanov transformation, there exists a probability $\widehat{\Q}^T$, locally equivalent to $\widehat \P$, and $\widehat{\Q}^T$-Brownian motions $\widehat{W}^{\widehat{\Q}^T}$ and  $\widehat {B}^{(1),\widehat{\Q}^T}$ such that \reff{CLE} is rewritten in $[0,T]$ as
\beqs
dZ_s\;=\; -\widehat{C}^*(s,Z_s)ds+ \sigma_L\widehat{\Pi}^*(s,Z_s)d\widehat W^{\widehat {\Q}^T}_s + K_\gamma Z_s d\widehat {B}^{(1),\widehat{\Q}^T}_s.
\enqs
By nonnegativity of $\widehat{C}^*$ and $Z^{t,z,*}$, the process $Z^{t,z,*}$ is a nonnegative $\widehat{\Q}^T$-supermartingale on $[t,\tau^T)$. It can be extended to a $\widehat{\Q}^T$-supermartingale  ($L^1(\widehat\Q^T)$-bounded)  on $[t,T]$ by setting $Z^{t,z,*}\equiv 0$ in $[\tau^T,T]$. Hence, by Doob's convergence Theorem (the usual proof for deterministic intervals - see e.g. Theorem 6.18 in \cite{Kallenberg} - can be adapted to our stochastic interval $[t,\tau^T)$), there exists a finite random variable $Z^{t,z,*}_{\tau^T}$ such that
$\lim_{s \nearrow \tau^T} Z^{t,z,*}_s \ = \ Z^{t,z,*}_{\tau^T}, \ \widehat{\Q}^T\mbox{-a.s.}.$
As $\widehat{\Q}^T$ $\sim$ $\widehat \P$, we also have
\beq \label{ineqQZeps}
\lim_{s \nearrow \tau^T} Z^{t,z,*}_s \ = \ Z^{t,z,*}_{\tau^T}, \;\;\;\;\;\;\;\ \widehat \P\mbox{-a.s.}.
\enq
Hence, \eqref{ineqQZeps} yields the desired extension on $\{\tau^T=T\}$. Consider now the set $\{\tau^T<T\}$.
In this set, $Z^{t,z,*}_{\tau^T_\eps} \in \{\eps, \eps^{-1}\}$, so, by \eqref{ineqQZeps}, necessarily $Z^{t,z,*}_{\tau^T} = 0$ $\P\mbox{-a.s.}$.
This clearly implies
$
\lim_{s \nearrow \tau} Z^{t,z,*}_s \;= \;0, \ \widehat \P - \mbox{a.s. on } \ \{\tau^T<T\}.
$
Hence, we can extend $Z^{t,z,*}$ to a solution defined over $[t,T]$ on $\{\tau^T<T\}$, by  setting
$ Z_s^{t,z,*}\ \equiv\ 0 \  \mbox{for}\  s\in [\tau^T,T].$

\emph{Uniqueness.}
The solution is  unique on the stochastic interval $[t,\tau^T)$, defined in the existence part. On the set $\{\tau^T<T\}$, when it reaches $0$, it must stay there, as it is a nonnegative $\widehat{\Q}^T$-supermartingale. Therefore, we have uniqueness on $[t,T]$ for all $T>t$, hence  on $[t,+\infty)$.
\hfill$\square$

\subsection{Verification theorem}

  \begin{Theorem}\label{th:ver}
Let  $Z^{t,z,*}$ be the unique nonnegative solution to \eqref{CLE}  and let $\widehat{C}^*, \widehat{\Pi}^*$ be the feedback maps defined in \eqref{def:feed}.
  Define the feedback strategies
\beq\label{strfeed}
\hat{c}^*_s\;:=\;\widehat{C}^*(s,Z^{t,z,*}_s), \ \ \ \hat{\pi}^*_s\;:=\;\Pi^*(s,Z^{t,z,*}_s), \ \ \ s\geq t.
\enq
Then $(\hat{c}^*,\hat{\pi}^*)\in \hat{\mathcal{A}}_{t}(z)$, and it is the unique optimal control  for \eqref{def_phi}.
\end{Theorem}
\textbf{Proof.}
\emph{Admissibility.}
As \eqref{CLE}  has a well-defined solution,  then $\hat{c}^*$ and $\hat{\pi}^*$ satisfy the required integrability conditions $\mbox{(h1)}''-\mbox{(h2)}''$.
On the other hand, by uniqueness of solutions to \eqref{CLE}, we must have
 $
{Z}^{t,x,\hat{c}^*,\hat{\pi}^*}=Z^{t,z,*}.
 $
 As $Z^{t,z,*}\geq 0$, we conclude $(\hat{c}^*,\hat{\pi}^*)\in \hat{\mathcal{A}}_{t}(z)$.

\emph{Optimality}. To prove optimality  we distinguish two cases\,: $z=0$ and $z>0$.

\emph{(i) Case $z=0$}.  Then $Z^{t,z,*}\equiv 0$, so also $(\hat{c}^*,\hat{\pi}^*)\equiv (0,0)$. On the other hand, $\hat{\mathcal{A}}_t(0)=\{(0,0)\}$, hence we conclude.

\emph{(ii) Case $z>0$}.
Let $\tau  :=
 \inf\{s\geq t \, | \, {Z}^{t,z,\hat{c}^*,\hat{\pi}^*}_s=0\}$, with the convention $\inf\emptyset =+\infty$.
Due to Theorem \ref{prop:reg},  we can apply  Dynkin's formula  to
$s\mapsto e^{-K_\lambda(s-t)} \Phi(s,Z^{t,z,\hat{c}^*,\hat{\pi}^*}_s)$ in  $[t,\tau\wedge T)$, for all $T>t$.
By Proposition \ref{prop:visc2} and Theorem \ref{prop:reg}, $\Phi$ solves \eqref{HJBphi*} in classical sense. Hence, using the definition of $(\hat{c}^*,\hat{\pi}^*)$ and arguing as in standard verification theorems we get
\begin{multline*}
\Phi(t,z) -  \widehat \E\left[e^{-K_\lambda((\tau\wedge T)-t)}\Phi(\tau\wedge T,Z_{\tau\wedge T}^{t,z,\hat{c}^*,\hat{\pi}^*})\right] \\
 =\; \widehat \E\left[ \int_t^{\tau\wedge T} e^{-K_\lambda (s-t)} \left( \frac{(\hat{c}^*_s)^p}{p} + \lambda K_V  f_\gamma(s,Z^{t,z,\hat{c}^*,\hat{\pi}^*}_s)\right) ds\right].
\end{multline*}
Splitting on the sets ${A}_T=\{\tau<T\}$ and $A_T^c=\{\tau\geq T\}$, we write
\begin{multline}\label{sd}
 \Phi(t,z) - \widehat \E\left[\mathbf{1}_{A_T} \ e^{-K_\lambda (\tau-t)}\Phi(\tau,Z^{t,z,\hat{c}^*,\hat{\pi}^*}_\tau)+\mathbf{1}_{A_T^c} \ e^{-K_\lambda (T-t)}\Phi(T,Z^{t,z,\hat{c}^*,\hat{\pi}^*}_T)\right] \\
 = \;\widehat \E\Bigg[\mathbf{1}_{A_T} \int_t^{\tau} e^{-K_\lambda (s-t)}\left( \frac{(\hat{c}^*_s)^p}{p} + \lambda K_V  f_\gamma(s,Z^{t,z,\hat{c}^*,\hat{\pi}^*}_s)\right) ds\\
 +\mathbf{1}_{A_T^c}  \int_t^{T} e^{-K_\lambda (s-t)} \left( \frac{(\hat{c}^*_s)^p}{p} + \lambda K_V  f_\gamma(s,Z^{t,z,\hat{c}^*,\hat{\pi}^*}_s)\right) ds\Bigg].
\end{multline}
Noting that $Z^{t,z,\hat{c}^*,\hat{\pi}^*}\equiv 0$ and $\hat{c}^*\equiv 0$, from $\tau$ on, using \eqref{bbb}  we get
\begin{multline*}
\widehat \E\left[\mathbf{1}_{A_T} \ \left(e^{-K_\lambda (\tau-t)}\Phi(\tau,Z^{t,z,\hat{c}^*,\hat{\pi}^*}_\tau)+ \int_t^{\tau} e^{-K_\lambda (s-t)}\left( \frac{(\hat{c}^*_s)^p}{p} + \lambda K_V  f_\gamma(s,Z^{t,z,\hat{c}^*,\hat{\pi}^*}_s)\right) ds\right)\right]\\
 =\ \ \widehat \E\Bigg[\mathbf{1}_{A_T} \ \int_t^{\infty} e^{-K_\lambda(s-t)}  \left( \frac{(\hat{c}^*_s)^p}{p} + \lambda K_V  f_\gamma(s,Z^{t,z,\hat{c}^*,\hat{\pi}^*}_s)\right)  ds\Bigg].
\end{multline*}
Hence,
moving the term corresponding to $\mathbf{1}_{A_T}$  to the right hand side in \eqref{sd}, and adding and subtracting $\widehat \E\Bigg[\mathbf{1}_{A_T^c}  \int_T^{+\infty} e^{-K_\lambda(s-t)} \left( \frac{(\hat{c}^*_s)^p}{p} + \lambda K_V  f_\gamma(s,Z^{t,z,\hat{c}^*,\hat{\pi}^*}_s)\right) ds\Bigg]$,  we get
\begin{multline}\label{sd1}
  \Phi(t,z) -  \widehat \E\left[\mathbf{1}_{A_T^c} \ e^{-K_\lambda (T-t)}\Phi(T,Z^{t,z,\hat{c}^*,\hat{\pi}^*}_T)\right]\\
 =\;\widehat  \E\Bigg[\int_t^{\infty} e^{-K_\lambda(s-t)}  \left( \frac{(\hat{c}^*_s)^p}{p} + \lambda K_V  f_\gamma(s,Z^{t,z,\hat{c}^*,\hat{\pi}^*}_s)\right)  ds\Bigg]\\
 -\widehat \E\Bigg[\mathbf{1}_{A_T^c}  \int_T^{+\infty} e^{-K_\lambda(s-t)} \left( \frac{(\hat{c}^*_s)^p}{p} + \lambda K_V  f_\gamma(s,Z^{t,z,\hat{c}^*,\hat{\pi}^*}_s)\right) ds\Bigg].
\end{multline}
Take  $T\rightarrow {\infty}$ in \eqref{sd1}. The second term of the left hand side goes to $0$ by dominated convergence, due to
 Assumption \ref{ass:beta}, \eqref{hhj1}, \eqref{newgrowth}; the second term of the right hand side goes to $0$ by monotone convergence. So we conclude.

\emph{Uniqueness.}  As $V$ is strictly concave  (it is of power form),  $f_\gamma(s,\cdot)$ is also strictly concave.
Let  $(\hat{c}^*_1,\hat{\pi}^*_1)$ and $(\hat{c}^*_2,\hat{\pi}^*_2)$ be optimal for \eqref{def_phi}.
Strict concavity of  $c\mapsto c^p/p$  yields $\hat{c}^*_1=\hat{c}^*_2$. Strict concavity of $f_\gamma (s,\cdot)$, for every $s\geq t$, yields $Z^{t,z,\hat{c}^*_1,\hat{\pi}^*_1}= Z^{t,z,\hat{c}^*_2,\hat{\pi}^*_2}$, from which $\hat{\pi}^*_1=\hat{\pi}^*_2$  follows.
\hfill\ep
 \subsection{Optimal policies for  the original problem}
First, we characterize the optimal policy in the illiquid asset.
\begin{Proposition}\label{prop:OI}
The  optimal allocation policy $\alpha_k^*$  in the illiquid asset at time $\tau_k$ is $\alpha_k^*= a^*(R_{\tau_k})$,
where $a^*(r)$ is the unique maximizer over $[0,r]$ of the function $\mapsto g(a;r)= a^p\Phi(0,\frac{r}{a}-1)$.
\end{Proposition}
\textbf{Proof.} As  $\Phi\geq 0$, $\Phi_z>0$, and $\Phi_{zz}<0$, a computation of $\partial^2 g/\partial a^2$ shows that $g(\cdot;r)$ is strictly concave in $(0,r)$, so it admits a unique maximizer in $[0,r]$.
The claim follows from  Proposition \ref{prop:conn}(3) and  \eqref{alphaopt}. \hfill$\square$\\

We prove
two important properties of the optimal allocation in the illiquid asset.
\begin{Proposition}\label{prop:z*}
The optimal rebalancing proportion $\frac{X^*_{\tau_k}}{\alpha_k^*}$ at the trading times of the illiquid asset $(\tau_k)_{k\in\N}$ is constant\,:
\beqs
\frac{X^*_{\tau_k}}{\alpha_k^*}\ =\ \frac{R^*_{\tau_k}-\alpha_k^*}{\alpha_k^*} &=& z^*\ :=\ \mbox{argmax}_{z\geq 0} \ \frac{\Phi(0,z)}{(1+z)^p},\ \ \ \ \forall k\in\N,
\enqs
where the value $z^*$ above is well defined, under the convention that $z^*=\infty$ if the supremum of $\frac{\Phi(0,z)}{(1+z)^p}$ is not attained (in this case,    there is no investment in the illiquid asset).

Consequently, at $(\tau_k)_{k\in\N}$, the optimal allocation proportions in liquid and illiquid assets over total wealth are also constant:
$\frac{\alpha_k^*}{R_{\tau_k}^*} =  \frac{1}{1+z^*}, $ $ \frac{X_{\tau_k}^*}{R_{\tau_k}^*} = \frac{z^*}{1+z^*},$ for every $k\in\N,$
with the conventions $\frac{1}{\infty}=0$, $\frac{\infty}{\infty}=1$.
\end{Proposition}
\textbf{Proof.} Consider
$h(z): =   \frac{\Phi(z)}{(1+z)^p},$  $ z\geq0.$
We have $h(\frac{r}{a}-1)=g(a;r)$ for $a\in (0,r]$, where $g$ is  defined in Proposition \ref{prop:OI}. As $g$ is continuous,  $\lim_{z\rightarrow +\infty} h(z)$ exists and is equal to $g(0)$. So, setting $h(+\infty):=g(0)$,
we consider  the diffeomorphism
$[0,r]\rightarrow [0,+\infty],  \ a\mapsto \frac{r}{a}-1,$
and note that $a$ maximizes $g$ over $[0,r]$ if and only if $z=  \frac{r}{a}-1$ maximizes $h$ over $[0,+\infty]$. The maximizer of $g$ over $[0,r]$ is unique, hence  the maximizer of $h$ over $[0,+\infty]$ is also unique. Calling it $z^*$, from the correspondence above,  we get $z^*= \frac{r-a^*(r)}{a^*(r)}$, where $a^*(r)$ is defined in Proposition \ref{prop:OI}.
%
\hfill\ep

\medskip
%
%
\begin{Proposition} \label{propCNS}
\begin{enumerate}\item[]
\item
$\alpha_0^*<r$ (if and only if $r>0$).
\item
$\alpha^*_0>0$ if and only if $\frac{b_I}{\sigma_I} > \frac{\rho b_L}{\sigma_L}$.\footnote{This condition is the same as  in the Merton (liquid) problem with two assets. The same result is obtained in \cite{APW} in the case of full observation.}
\end{enumerate}
\end{Proposition}

\ni{\bf Proof.}
1. If $r=0$, then  $\alpha_0^*=0$, due to the state constraint. Let $r>0$ and assume, by contradiction, that $\alpha^*_0=r$. This would yield $z^*=0$ in Proposition \ref{prop:z*}, hence $\alpha_k^*=R,_{\tau_k}$ for all $k\in\mathbb{N}$. We should conclude, by the state constraint, that  $c_t^*\equiv 0$. But this strategy cannot be optimal, as $V(r)>0$.

2.
\emph{Necessity}.
Consider the Merton problem described in Remark \ref{remMerton}
%
and call its  value function $V^{M,2}$.
 The optimal investment proportions in $L$ and $I$ for this problem are
\beqs
(u_L^*,u_I^*) &=& \argmax_{u_L \in \R,\, u_I \in [0,1]} \left\{p (u_L b_L + u_I b_I) - \frac{p(1-p)}{2} ( u_L^2 \sigma_L^2 + u_I^2 \sigma_I^2 + 2 \rho u_L u_I \sigma_L \sigma_I)\right\}
\enqs
Taking first the supremum on  $u_L$, one sees that
$u_I^*=0$ if and only if $b_I$ $\leq$ $\frac{\rho b_L \sigma_I}{\sigma_L}$. In this case, denoting by $V^{M,1}$ the value function for an agent investing only in $L$, we have $V^{M,2} = V^{M,1}$. As $V^{M,1}$ $\leq$ $V$ $\leq$ $V^{M,2}$, we obtain $V$ $=$ $V^{M,1}$, and the optimal strategy for our original problem never invests in the illiquid asset $I$.

\vspace{2mm}
\emph{Sufficiency.}
Assume $\frac{b_I}{\sigma_I} > \frac{\rho b_L}{\sigma_L}$ and set $h(a) := \widehat V(0,1-a,a)$, $a\in[0,1]$. By \eqref{VVV}, it suffices to show that $h'(0^+)>0$.
We have
\beqs
h'(0^+) &=& \lim_{\eta \to 0} \,\frac{\widehat V(0, 1- \eta, \eta) - \widehat V(0,1,0)}{\eta} \\
&=& \lim_{\eta \to 0} \,\frac{(1-\eta)^p}{\eta} \Big(\widehat V(0, 1, \frac{\eta}{1-\eta}) - \widehat V(0,1+ \frac{\eta}{1-\eta},0)\Big) \\
&=&\left(\lim_{\eta\rightarrow 0}\ (1-\eta)^{p-1}\right)\ \left( \lim_{\delta \to 0} \,\frac{1}{\delta} \big( \widehat V(0,1,\delta) - \widehat V(0,1+\delta,0)\big)\right)\\
&=& \widehat V_y(0,1,0^+) - \widehat V_x(0,1^+,0).
\enqs
We will show that the latter is strictly positive.
 Consider the auxiliary problem with initial data $(t,x,y)=(0,1,0)$.
In this case the problem is the  Merton type problem \eqref{degcon}, with  value function $\widehat V
 (0,x,0)=  K_0 x^p$, so
 $\widehat V_x(0,1^+,0)=  pK_0$.
By solving the HJB equation for this problem, one finds $K_0$ as unique positive solution to
 \beq \label{eqK0}
 \Big(\beta + \lambda - \frac{p b_L^2}{2 (1-p) \sigma_L^2}\Big) K_0 - (1-p) p^{-\frac{1}{1-p}} K_0^{- \frac{p}{1-p}} & =&  \lambda {K_V},
\enq
and the corresponding optimal wealth process is
\beq \label{eqX0}
d \tilde{X}^*_t = -{c}^*_t dt + {\pi}^*_t \frac{dL_t}{L_t},
\enq
where
\beq\label{cpi}
{c}^*_t\ =\  p^{- \frac{1}{1-p}} K_0^{- \frac{1}{1-p}} \ \tilde{X}^*_t , \;\;\;\; \ \ \ \ {\pi}^*_t \ = \ \frac{b_L}{\sigma_L^2(1-p)}\ \tilde{X}^*_t .
\enq
Considering an agent with initial wealth $(1,\delta)$, who chooses the same investment/consumption strategy, we
get
\beqs
\widehat{V}(0,1,\delta)&\geq& \mathbb{E}\Big[\int_0^{\infty}e^{-(\beta+\lambda)t} \left(\frac{({c}_t^*)^p}{p}+\lambda G_\gamma[V](t,\tilde{X}^*_t,\tilde{Y}_t^{0,\delta})\right)dt\Big].
\enqs
Therefore
\beq\label{pllsk}
\frac{\widehat{V}(0,1,\delta)-\widehat{V}(0,1,0)}{\delta}&\geq&\frac{\lambda}{\delta}\; \E\left[\int_0^{\infty}e^{-(\beta+\lambda)t}\big(G[V](t,\tilde{X}^*_t,\tilde{Y}_t^{0,\delta})-G[V](t,\tilde{X}^*_t,0)\big) dt\right]\nonumber\\
&=&\lambda{K_V}\; \int_0^{\infty}e^{-(\beta+\lambda)t}\ \E \left[\frac{(\tilde{X}^*_t+\tilde{Y}_t^{0,\delta}J_t)^p-(\tilde{X}^*_t)^p}{\delta}\right] dt.
\enq
Let $\delta\rightarrow 0$ in \eqref{pllsk}. Applying Fatou's Lemma and observing that $\tilde{Y}^{0,\delta}J_t=\delta I_t$,
\beqs
\widehat V
_y(0,1,0^+) &\geq& p {K_V} \lambda \int_0^\infty e^{- (\beta+\lambda)t} \E\left[ (\tilde{X}^*_t)^{p-1} I_t\right]dt \\
&=& p {K_V} \lambda \int_0^\infty \exp\Big( - \big(\lambda \frac{{K_V}}{K_0} - (b_I - \frac{\rho b_L \sigma_I}{\sigma_L})\big) t \Big) dt\\
&>& p  K_0 \;\;\; = \;\;\;\widehat V
_x(0,1^+,0),
\enqs
where the  middle equality  uses \eqref{LL}, \eqref{II},  \reff{eqK0}, \reff{eqX0} and \eqref{cpi}, and  the strict inequality uses $\frac{b_I}{\sigma_I} > \frac{\rho b_L}{\sigma_L}$.
\hfill$\square$

\medskip
When the two assets are uncorrelated, the result above says, in particular, that there is investment in the illiquid asset even if the Sharpe ratio of the liquid asset is higher than that of the illiquid asset.
Let us now deal with the optimal consumption and  investment in the liquid asset.
Set,
for any $x \geq 0$, $y>0$,\footnote{For the case $y=0$, see the discussion after \eqref{degcon}.}
 $$C^*(s,x,y)\ := \ y \ \widehat{C}^*\Big(s,\frac{x}{y}\Big), \ \ \ \ \ \Pi^*(s,x,y)\ :=\ y \  \widehat{\Pi}^*\Big(s,\frac{x}{y}\Big)  + \frac{\rho \sigma_I}{\sigma_L} \frac{x}{y},$$
  and set $\tilde{X}^{*,t,x}:=Z^{*,t, x/y}\tilde{Y}^{t,y}$.
From the above results, we get the following theorem.\,\footnote{We make the assumption $\frac{b_I}{\sigma_I} > \frac{\rho b_L}{\sigma_L}$ to make the problem meaningful in view of Proposition \ref{propCNS} (see again the discussion after \eqref{degcon}). In this case $z^*<\infty$.}
 \begin{Theorem}\label{Prop:opt}
  Let $\frac{b_I}{\sigma_I} > \frac{\rho b_L}{\sigma_L}$.
The unique optimal control $(\alpha^*,c^*,\pi^*)$ for \eqref{eq:optprob} is
$$
\begin{cases}
\alpha_k^* \  =  \displaystyle{\frac{R_{\tau_k}}{1+z^*}}, \ \ \ \ \ \ \   k\in\N,\\\\
c^*_s\ =\ C^*(s-\tau_k,\tilde{X}_s^{*,\tau_k,R_{\tau_k}-\alpha_k^*}, \tilde{Y}^{\tau_k,\alpha^*_k}_s), \ \ \ \ \ \ \ s\in[\tau_k,\tau_{k+1}), \ \ k\in\N,\\\\
\pi^*_s \ =\ \Pi^*(s-\tau_k,\tilde{X}_s^{*,\tau_k,R_{\tau_k}-\alpha_k^*,}, \tilde{Y}^{\tau_k,\alpha^*_k}_s), \ \ \ \ \ \ \  s\in[\tau_k,\tau_{k+1}), \ \ k\in\N.
\end{cases}
$$
\end{Theorem}
\textbf{Proof.}
The expression for $\alpha_k^*$ follows from Proposition \ref{prop:z*}.
The proof that $(c^*,\pi^*)$ is the unique optimal control in each random interval $(\tau_k, \tau_{k+1})$ {follows by doing again, but in the opposite direction, the transformations  leading to the equivalence between the original problem \eqref{eq:optprob} and the transformed one \eqref{def_phi} (via the first transformation \eqref{defvhat1}). Precisely, using Proposition \ref{prop:conn}, Theorem \ref{th:ver},  Proposition \ref{propDPP}, and  the Markov nature of our problem,  one shows, by induction, that
$ V(r) =  \E\left[\int_0^{\tau_k} e^{- \beta s} U(c^*_s) ds\right]  + \E\left[e^{- \beta \tau_k} V\left(R^*_{\tau_k}\right) \right],$  for all $k\in\N$.
Then, \eqref{Phi_0} implies that the second term on the right hand side  goes to $0$ for $k \to \infty$, hence
 $(\alpha^*,c^*,\pi^*)$ is optimal.}
\hfill$\square$
%
%
%

\subsection{Numerical approximations}\label{Sec:num}

\setcounter{equation}{0}

This subsection  presents an iterative scheme to approximate ${K_V}$ and $\Phi$. The procedure is illustrated more extensively, in the case $\gamma=0$,  in \cite{FGtec, gasphd}.
 Because of the nonlocal term $\mathcal{H}_0[\varphi]$ in \eqref{HJBphi**}, we cannot approximate  the value function $\Phi$ directly as a viscosity solution of a PDE, but  need to define an iterative scheme. Fixing $T>0$ and starting with
$ K_V^{0,T} := 0$,
  define, inductively on $n\in\N$, the sequence $(K_V^{n,T},\Phi^{n,T})$ as follows.
\begin{itemize}
\item[-] Given $n\in\N$ and $K_V^{n,T}$, let $\Phi^{n,T}$ on $\R_+^2$ be the unique constrained viscosity solution on $[0,T]\times \R_+$ to
\beq\label{PDEn}
- \Phi^{n,T}_t + {K_\lambda}  \Phi^{n,T} -  \lambda  K_V^{n,T} f_\gamma(t,z)
 -\sup_{\hat{c}\geq 0,\hat{\pi}\in \R}{H}^0_{cv}(z,\Phi^{n,T},\Phi^{n,T}_z,\Phi^{n,T}_{zz}; \hat{c},\hat{\pi})=0,
\enq
with boundary and  terminal conditions
\beq \label{Bndryvn}
{\Phi}^{n,T}(t,0)& =& K_V^{n,T} \int_t^T e^{-K_\lambda(s-t)}\lambda f_\gamma(s,0) ds, \ \ \ \ t\in[0,T],
\enq
\beq \label{Bndryvnt}
{\Phi}^{n,T}(T,z)& =& 0, \ \ \ \ z\geq 0.
\enq
\item[-]Given $n\in\N$ and $\Phi^{n,T}$, let $K_V^{n+1,T}:= \mathcal{H}_0 [\Phi^{n,T}]$.
\end{itemize}
Then one proves, following  \cite[Ch.\,3, Sec.\,6]{gasphd}, the following estimate ensuring the convergence of our scheme when $T\rightarrow\infty$ and $n\rightarrow \infty$.
\begin{Proposition}\label{prop:estimate}
Let
$
\delta := \frac{\lambda}{\lambda + \beta - k_p}.
$
 For all $(t,z)$ in $[0,T] \times \R_+$,
\beqs
\left|(\Phi - \Phi^{n,T})(t,z)\right| &\leq& C_0 e^{k_{J,p} t} (1+z)^p \Big(\delta^n + \frac{e^{-(\lambda + \beta - k_p)T}}{1-\delta} + e^{-(\lambda + \beta - k_p)(T-t)}\Big),
\enqs
where $C_0=V^{(p)}_{Mert}(1)$ (see Remark \ref{remMerton}).
\end{Proposition}
Proposition \ref{prop:estimate}
provides a rate of convergence sensitive to the value of $\lambda$\,:
\begin{itemize}
\vspace{-.1cm}
	\item[-] the larger is $\lambda$, the slower is the convergence in $n$;
	\vspace{-.3cm}
	\item[-] the smaller is $\lambda$, the slower is the convergence in $T$.
\end{itemize}

\section{Discussion}
This section provides and discusses some numerical experiments
performed, in the case of power utility, by means of the iterative approximation procedure  described in Subsection \ref{Sec:num}.
The  discussion  is limited to  key features, in order to show how our methodology can be applied.
We choose  the parameters
\beqs
\beta \ = \ 0.2, \;\;\;\;\;\;p\ =\ 0.5,\;\;\;\;\;\;b_L \ = \ 0.15, \;\;\;\;\;\;\sigma_L \ = \ 1, \;\;\;\;\;\; b_I\ =\ 0.2, \;\;\;\;\;\;\sigma_I \ =\  1.
\enqs
With these values, Assumption \ref{ass:beta} is satisfied for every $\rho\in(-1,1)$.
It is reasonable to let the illiquid asset have a higher Sharpe ratio than the liquid one. This is economically intuitive and  ensures that, for every value of the correlation $\rho$, it is always  optimal to invest in the illiquid asset (Proposition \ref{propCNS}(2)).
We solved the PDE \eqref{PDEn} using an explicit finite-difference scheme, 
 after the change of variable
$\R_+ \rightarrow  [0,1), \  z \mapsto \tilde{z} = \frac{z}{z+1},$
inducing a  corresponding transformation
$\Phi\mapsto \tilde{\Phi},$  to work with the bounded domain $[0,1)$.
We fixed beforehand $T$ between $1$ and $10$, depending on $\lambda$ according to what we said at the end of  Subsection \ref{Sec:num}\,: for the extreme cases $\lambda= 1$ and $\lambda =50$, we took, respectively, $T=10$ and  $T=1$, as with these choices the terms involving $T$ and $T-t$ in Proposition \ref{prop:estimate} become reasonably negligible (resp., of order $10^{-5}$ and $10^{-22}$).
We used a uniform grid on $[0,T] \times [0,1]$ with time step length $5\cdot10^{-4}$ and space step length $0.02$.
The numbers $f_\gamma(t,\tilde{z})$ were computed beforehand at each point of the grid, using an $L^2$-optimal quantization grid for the gaussian law with $N=5000$ points. 
Finally, almost all the numerical tests below were performed for $t=0$, so Proposition \ref{prop:estimate} was used with $t=0$ to estimate the error. The only numerical tests performed with $t>0$ are the ones regarding the optimal consumption and investment in the liquid asset  in Subsection \ref{sub:oppol}, where $\lambda =5$ and $t=1$; in this case,  $T=5$ in order to control the error.

\subsection{Value function and cost of illiquidity}
To study   the cost of  illiquidity, we consider quantities related to the value function $V$.
The first  measure of illiquidity cost  is  the difference between the value functions corresponding to different values of $\lambda$.\footnote{We stress that
one cannot always expect convergence to Merton's unconstrained solution  when $\lambda\rightarrow\infty$. The presence of illiquidity - even in the case of high $\lambda$ (high trading frequency)  - induces a constraint on the investment strategies in $I$, in order to satisfy  the state constraint.  This fact can produce a gap between the fully liquid case and the limit of the illiquid one. In general, the limit case for $\lambda\rightarrow\infty$ corresponds to the \emph{constrained} fully liquid Merton problem, i.e. the Merton problem with the constraint that the investment  in $I$ does not admit borrowing or short selling. When the optimal solution of the unconstrained Merton problem satisfies this constraint, the constrained and unconstrained Merton problems are  equivalent\,: hence, in the latter case,  we have  convergence to the (unconstrained) Merton solution for $\lambda\rightarrow\infty$.}
As $V(r)=r^pV(1)$, we study $V(1)$.
 Table \ref{tbl_V}   reports  it for different values of $\lambda$ and for $\gamma=0$ or $\gamma=1$.

\begin{table}[h]
\centering
\begin{tabular}{|c|cccc|c|}
\hline
$\lambda$ & 1 & 5 & 10 & 50 & Constr./Unconstr. Merton \\
\hline
$\gamma=0$ & 1.66755 & 1.70493 & 1.71257 & 1.71945 & 1.72133 \\
\hline
$\gamma=1$ & 1.67179 & 1.71121 & 1.71656 & 1.72036 & 1.72133 \\
\hline
\end{tabular}
\caption{\small{$V(1)$ for various $\gamma$, $\lambda$, and fixed $\rho=0$.}}
\label{tbl_V}
\end{table}

%
%

Another  measure  of illiquidity cost  is given by - see \cite{phatan08} -   the extra amount of initial wealth $e(r)$ needed to reach the same level of expected utility as an investor without trading restrictions and with the same  initial capital $r$. Hence, it corresponds to the solution to
the equation
$V(r + e(r))	=  V_M (r),$
where $V_M$ is the value function of the corresponding unconstrained Merton problem.
As  $e(r)$ is proportional to $r$, we  study $e(1)$.
Table \ref{tbl_e} reports $e(1)$   for different values of $\lambda$ and for $\gamma=0$ or $\gamma=1$.

\begin{table}[h]
\centering
\begin{tabular}{|c|cccc|}
\hline
$\lambda$ & 1 & 5 & 10 & 50\\
\hline
$\gamma=0$ & 0.066 & 0.0193 & 0.0103 & 0.00218 \\
\hline
$\gamma=1$ & 0.060 & 0.0119 & 0.0056 & 0.00112 \\
\hline
\end{tabular}
\caption{\footnotesize{$e(1)$ for various $\gamma$, $\lambda$, and fixed $\rho =0$.}}
\label{tbl_e}
\end{table}
Concerning the impact of the observation parameter $\gamma$, we observe that:
\begin{enumerate}
\item The impact of $\gamma$ on the absolute cost of liquidity, measured as $V(1)$ or as $e(1)$, is not high, but shows a peak for intermediate values of $\lambda$.
 This is expected: when $\lambda$ is low, the illiquid asset is very rarely traded, so it is less useful to have information on it;  when $\lambda$ is  high,  the discrete information on $I$ at trading dates is  frequently updated, so the continuous part of the information on $I$ is less relevant.
 \vspace{-.2cm}
\item {The relative impact of  $\gamma$ on $e(1)$, i.e. the quantity $\frac{e^{\gamma=0}(1)-e^{\gamma=1}(1)}{e^{\gamma=0}(1)}$,  for   $\lambda\geq 5$ is of the order of $50\%$.}
\end{enumerate}

  \subsection{Optimal policies}\label{sub:oppol}
\begin{figure}[h]
\include{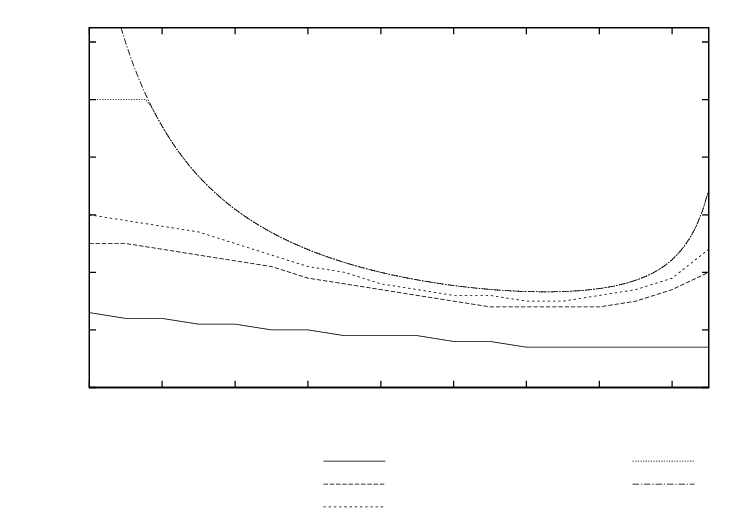}
\vspace{-1cm}
\caption{\footnotesize{Optimal investment proportion $\hat{z}$  in the illiquid asset, as function of $\rho$, for $\gamma=0$}.}
\label{figA}
\end{figure}

 Again, without loss of generality, we assume that $r=1$.
Figure \ref{figA}   represents the optimal allocation policy in the illiquid asset, expressed as a proportion of wealth - i.e. the quantity
$\hat{z}:=\frac{1}{1+z^*}$ - and as function of  $\rho$, for fixed $\gamma=0$. The different lines correspond to  different values of $\lambda$. When $\lambda$ is low, $\hat{z}$ is close to $0$; when $\lambda$ is  high, $\hat{z}$ is close to the corresponding value in the constrained Merton problem.
For increasing values of $\lambda$, the  graphs lie between these two extreme cases,  increasing with it.


%

The impact of the observation parameter on the optimal allocation proportion in the illiquid asset is shown in Table \ref{tbl-z^*} for different values of $\lambda$, considering the extreme cases $\gamma=0$ and $\gamma=1$, for fixed $\rho =0$. The agent  invests more in the illiquid asset if she/he can observe it continuously.
 The impact of $\gamma$ is negligible in the extreme cases $\lambda=1$ and $\lambda=50$; of the order of $6\%$, when $\lambda=3, 5, 10$.

\begin{table}[h]
\centering
\begin{tabular}{|c|ccccc|c|}
\hline
& $\lambda=1$ & $\lambda=3$ & $\lambda= 5$ & $\lambda=10$ & $\lambda=50$ & Constr./Unconstr. Merton \\
\hline
$\gamma=0$ & 0.18 & 0.3 & 0.34 & 0.36 & 0.4 & 0.4 \\
$\gamma=1$ & 0.18 & 0.32 & 0.36 & 0.38 & 0.4  & 0.4\\
\hline
\end{tabular}
\caption{\small{Optimal investment (proportion over the wealth) in the illiquid asset,   for various $\gamma$, $\lambda$ and fixed $\rho =0$.}}
\label{tbl-z^*}
\end{table}

\vspace{3mm}

Let us analyze  the feedback maps $C^*$, $\Pi^*$.
We need to be careful comparing these maps for different values of  $\gamma$, when $t>0$. Indeed, they are defined on  $y$, which refers to the stochastic process $\tilde{Y}$. But $\tilde{Y}$  depends on  $\gamma$, so the feedback maps ${C}^*, {\Pi}^*$ do not read the same input for different values of  $\gamma$.
 To overcome this problem, one would need  to perform Monte-Carlo simulations to study the distributions of the optimal  strategies.\footnote{As our auxiliary control problem is not autonomous, we cannot look at the stationary distribution as in \cite{APW}.}
However, such an approach would be numerically intensive; for simplicity, we  consider the dependence of the feedback maps on the extra observation
$B^{(1)}$, for different values of $\gamma$, which still enables us  to illustrate the effect of partial observation.
As we are mainly interested  in the impact of $\gamma$ on the strategies,  we fix the other parameters, taking $\lambda=5$ and $\rho=0$.
We consider an agent who, at time $t=1$, owns a liquid wealth  $\tilde{X}_1=0.5$,  having invested $\alpha_0=\tilde{Y}_0=0.5$ in illiquid wealth at time $t=0$ (assuming that  $\tau_1$ has not occurred yet). We plot the optimal consumption   and  investment in the liquid asset  as  functions of the additional information $B^{(1)}_1$, which determines, together with $\gamma$, the value of $\tilde{Y}_1$.\footnote{This is true as $\rho=0$, so that $\tilde Y$ does not depend on $W$.}
To be more explicit, we  compute $C^*(1,\tilde{X}_1,\tilde{Y}_1)$,  $\Pi^*(1,\tilde{X}_1,\tilde{Y}_1)$ as functions  of $B_1^{(1)}$ and $\gamma$. As, from \eqref{XY2t}, 
$
\tilde{Y}_1 =
\tilde{Y}_0\
\exp\left(b_Y-\frac{(1-\rho^2)\gamma^2}{2}+\sqrt{1-\rho^2} \ \gamma B^{(1)}_1\right),
$
we get the function to plot by substitution.

\begin{figure}[h!]
\include{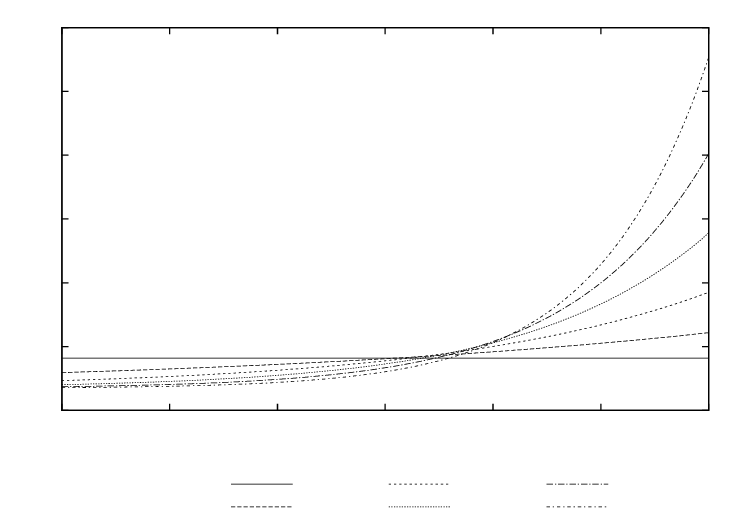}
\caption{\footnotesize{Optimal consumption rate, as function of $B_1^{(1)}$, for various $\gamma$ (setting $\lambda=5$,  $\rho =0$, $\tilde{X}_1=.5$, $\tilde{Y}_0=.5$).}}
\label{fig-ct}
\end{figure}


Figure \ref{fig-ct} graphs the consumption rate\,: it is increasing in $B^{(1)}$, which illustrates the unsurprising fact that, when the agent knows that her/his illiquid investment is doing well,  she/he should consume more. This effect is considerably stronger, when information about the illiquid asset increases ($\gamma$ close to 1). {The impact of the observation on the investment in the liquid asset can  be studied in a similar way. We omit the resulting graph, as it has the same qualitative features as Figure \ref{fig-ct} (when the agent knows her/his illiquid investment is doing well, she/he can take more risk with her/his investment in the other risky asset).}

{\subsection{Conclusions}
From the analysis  performed, we can extract the following behaviour with respect to the relevant parameters $\gamma, \lambda$.
\vspace{-.3cm}

\begin{enumerate}
\item The impact of $\gamma$ on the cost of illiquidity and on the investment in the illiquid asset  is negligible for low and high value of $\lambda$, and  has a peak for intermediate values of $\lambda$.
\vspace{-.2cm}
\item The investment in the illiquid asset is increasing in $\lambda$ and $\gamma$.
\vspace{-.2cm}
\item The consumption and the investment in the liquid asset are very sensitive with respect to  $\gamma$ for intermediate values of $\lambda$.\end{enumerate}}

\appendix
\section{Appendix}
\subsection{Technical results}
\setcounter{equation}{0}
\begin{Lemma} \label{lemBeforetau1}
Given $r \geq 0$, for any $(c,\pi,\alpha)$ $\in$ $\Ac(r)$, there exists $(c^0,\pi^0)$ $\in$ $\Ac_0(r-\alpha_0)$ such that
$
(c,\pi) \mathbf{1}_{\{t \leq \tau_1\}} = (c^0,\pi^0) \mathbf{1}_{\{t \leq \tau_1\}} \ (P \otimes ds - \mbox{a.e.}).
$
\end{Lemma}

{\bf Proof.} Using the definition of $\mathbb{G}$, by a monotone class argument, for each   $(\Gc_t)_{t\geq 0}$-predictable process $(c,\pi)$, we can find a $(\mathcal{W}_t\vee \mathcal{B}^{(1)}_t)_{t\geq 0}$-predictable  process  $(c^0,\pi^0)$ satisfying $(c,\pi) \mathbf{1}_{\{t \leq \tau_1\}} = (c^0,\pi^0) \mathbf{1}_{\{t \leq \tau_1\}}.
$ The  fact that  $(c,\pi,\alpha)$ $\in$ $\Ac(r)$ implies $(c^0,\pi^0)\in\Ac_0(r-\alpha_0)$ is straightforward. \ep

\begin{Proposition}\label{prop:G}
\begin{itemize}
\item[]
\item[(i)] $G_\gamma$ is well defined on the set of measurable functions with at most linear growth.
\vspace{-.2cm}
\item[(ii)] $G_\gamma$ is linear and positive, in the sense that it maps positive functions to positive ones. As a consequence, $G_\gamma$ is increasing, in the sense that
$\phi\ \leq \ \psi \ \Rightarrow  G_\gamma[\phi]\ \leq \ G_\gamma[\psi].$
\vspace{-.2cm}
\item[(iii)] $G_\gamma$ maps increasing functions to functions increasing with respect to both $x$ and $y$.
\vspace{-.2cm}
\item[(iv)] $G_\gamma$ maps concave functions to functions concave with respect to   $(x,y)$.
\vspace{-.2cm}
\item[(v)]  If $\psi(r)=r^p,$ $p\in (0,1)$, then ($k_{J,p}$ is defined in \eqref{kjp})
\begin{equation}\label{eqGgrowthG1}
G_\gamma[\psi](t,\xi x,\xi y)\ =\  \xi ^p G_\gamma[\psi](t,x, y), \ \ \ \forall t\geq 0, \ \forall (x,y)\in \R_+^2,  \ \forall  \xi\geq 0.
\end{equation}
\begin{equation}\label{eqGgrowthG}
0\ \leq\  G_\gamma[\psi](t,x,y)\ \leq\  e^{k_{J,p} t}(x+y)^p, \ \ \ \forall t\geq 0, \ \forall(x,y)\in\R^2_+.
\end{equation}
\item[(vi)]Let $p \in(0,1]$, and let $\psi$ be a $p$-H\"older continuous function on $\R_+$. Then, there exists some constant $C \geq 0$ such that, for all $t\geq 0$, $x,x',y,y'>0$, and $0\leq h \leq 1$,
\beq
|G_\gamma[\psi](t,x,y) - G_\gamma[\psi](t,x',y)| &\leq& C |x-x'|^p, \label{HoldGx} \\
|G_\gamma[\psi](t,x,y) - G_\gamma[\psi](t,x,y')|&\leq& C e^{k_{J,p} t} |y-y'|^p, \label{HoldGy}  \\
|G_\gamma[\psi](t,x,y) - G_\gamma[\psi](t+h,x,y)| &\leq& C_1 e^{k_{J,p} t} y^p h^{p/2}, \label{HoldGt}
\enq
\end{itemize}
\end{Proposition}
\textbf{Proof.} It parallels \cite{FGtec}, where the statement is proved in the case $\gamma=0$. \hfill\ep

\begin{Lemma} \label{lemGrowthXY}
Let $p\in(0,1)$ and let $k_{L,Y,p}$, $k_{J,p}$ be defined as in \eqref{klyp}-\eqref{kjp}. For every $(t,x,y)\in\R_+^3$ and $(c,\pi)$ $\in$ $\Ac_t(x)$,  we have
\beq \label{ineqGrowthXY}
\E \left[(\tilde{X}_s^{t,x,{c},{\pi}}+\tilde{Y}_s^{t,y})^p\right] &\leq& e^{k_{L,Y,p} (s-t)} (x+y)^p, \ \ \ \forall s\geq t.
\enq
In particular, combining \eqref{ineqGrowthXY} with Proposition \ref{prop:G}(v) and denoting $\varphi(r):=r^p$, we have
\beq \label{ineqGrowthGXY}
\E \left[G_\gamma[\varphi](s,\tilde{X}_s^{t,x,{c},{\pi}},\tilde{Y}_s^{t,y})\right] &\leq& e^{k_{J,p} t} e^{(k_{L,Y,p}+k_{J,p}) (s-t)}(x+y)^p.
\enq
\end{Lemma}
\textbf{Proof.} It parallels \cite{FGtec}, where the statement is proved in the case $\gamma=0$. \hfill\ep
\begin{Lemma} \label{lemSplitting}
Set
\beqs
f(u_L,u_I)& :=&p (u_L b_L + u_I b_I) - \frac{p(1-p)}{2} ( u_L^2 \sigma_L^2 + u_I^2 \sigma_I^2 + 2 \rho u_L u_I \sigma_L \sigma_I).
\enqs
Recalling \eqref{kp}, we have
$
k_{p} = \sup_{u_L\in \R ,u_I \in [0,1]} f(u_L,u_I)
$.
For any $b'_Y$, $b'_J$ such that $b'_Y+b'_J=b_I$, define
\beqs
f_{b_Y'}(u_L,u_Y):= \ p (u_L b_L + u_Y b'_Y)- \frac{p(1-p)}{2} ( u_L^2 \sigma_L^2 + u_Y^2 \sigma_I^2(\rho^2 + \gamma^2(1-\rho^2)) + 2 \rho u_L u_Y \sigma_L \sigma_I),
\enqs
\beqs
f_{b_J'}(u_J)&:=& p b'_J u_J - \frac{p(1-p)}{2} \sigma_I^2 (1-\rho^2)(1-\gamma^2) u_J^2,
\enqs
and
$
k'_{L,Y,p}:=\sup_{u_L\in \R ,u_{Y} \in [0,1]} f_{b_Y'}(u_L,u_Y)$, $
k'_{J,p} := f_{b_J'}(u_J).$
Then $k_p \leq k'_{L,Y,p} + k'_{J,p}$ and this inequality is an equality if  we choose
\beq \label{eqSplt}
b'_Y&=&\gamma^2 b_I+ (1- \gamma^2) \frac{b_L \rho \sigma_I}{\sigma_L}.
\enq
\end{Lemma}
\textbf{Proof.}
As $f_{b_Y'}(u_L,u_I)+f_{b_J'}(u_I)=f(u_L,u_I)$,
by definition of $k_p, k'_{L,Y,p},  k'_{J,p}$, we have
\beqs
k_p \ \ = \ \sup_{u_L\in \R ,u_I \in [0,1]} \big( f_{b_Y'}(u_L,u_I) + f_{b_J'}(u_I) \big)\  \ \leq\ \  k'_{L,Y,p} + k'_{J,p}.
\enqs
The maximizers of $f,f_{b_Y'}, f_{b_J'}$ always exist, so  the inequality above becomes an equality if and only if there exist a maximizer $(u_L^*,u_{Y}^*)$ of $f_{b_Y'}$ and a maximizer
$u_J^*$ of $f_{b_J'}$ such that $u_Y^*=u_J^*$.
If $\gamma$ $\in$ $(0,1)$, by strict convexity of $f'_{b_Y'}$ and $f_{b_J'}$, these maximizers are unique and can be computed explicitly with the first-order conditions, as
\beqs
u_J^* \ =\ \mbox{Proj}_{[0,1]}\left(\frac{b'_J}{(1-p) \sigma_I^2 (1-\rho^2) (1-\gamma^2)}\right),\ \ \
u_Y^*\ = \ \mbox{Proj}_{[0,1]}\left(\frac{b'_Y - \frac{b_L\rho \sigma_I}{\sigma_L}}{(1-p) \sigma_I^2 (1-\rho^2) \gamma^2}\right),
\enqs
As  $b'_Y+b'_J=b_I$, \reff{eqSplt} can be rewritten as
$
\frac{b'_J}{(1-\gamma^2)} = \frac{1}{\gamma^2}\left(b'_Y - \frac{b_L\rho \sigma_I}{\sigma_L}\right),
$
which implies $u_J^* = u_Y^*$.
To conclude, it remains to notice that, under \reff{eqSplt}, for $\gamma =0$ (respectively, $\gamma=1$), the function $f_{b_Y'}$ does not depend on $u_Y$ (respectively, the function  $f_{b_J'}$ does not depend on $u_J$), so  we can  choose $u_Y^*=u_J^*$.
 \hfill\ep\\\\
\noindent
Given $(\bar{t},\bar{z})\in\R^+\times (0,+\infty)$ and $\varepsilon\in(0,\bar{z})$,  we denote
\beq\label{D_epsilon}
D_\varepsilon(\bar{t},\bar{z})&:=&[\bar{t}, \bar{t}+\varepsilon)\times (\bar{z}-\varepsilon, \bar{z}+\varepsilon)\subset \R^+\times (0,+\infty).
\enq
\begin{Lemma}\label{lemma:sub}
Let $(\bar{t},\bar{z})\in\R^+\times (0,+\infty)$ and $\varepsilon\in(0,\bar{z})$.
\begin{enumerate}
\item
There exists $N_\varepsilon>0$ such that
\beq\label{phit}
\limsup_{h\rightarrow 0^+} \ \left| \frac{\Phi(t+h,z)-\Phi(t,z)}{h}\right|&\leq& N_\varepsilon, \ \ \ \ \forall \ (t,z)\in D_\varepsilon(\bar{t},\bar{z}).
\enq
\item $\Phi(t,\cdot)\in C^1((\bar{z}-\varepsilon,\bar{z}+\varepsilon);\R)$, for every $t\in [\bar t,\bar{t}+\varepsilon)$, and there exist $m_\varepsilon,M_\varepsilon>0$ such that
\beq\label{phiz}
m_\varepsilon&\leq & \Phi_z (t,z) \ \leq\  M_\varepsilon, \ \ \ \forall (t,z)\in D_\varepsilon(\bar{t},\bar{z}).
\enq
\item $\Phi(t,\cdot)$ is twice differentiable a.e. in $(\bar{z}-\varepsilon,\bar{z}+\varepsilon)$, for every $t\in [\bar t,\bar{t}+\varepsilon)$. Moreover, denoting by $\mathcal{O}_t^\varepsilon\subset (\bar{z}-\varepsilon,\bar{z}+\varepsilon)$  the set where  $\Phi(t,\cdot)$ is twice differentiable, there exists  $\delta_\varepsilon>0$ such that
\beq\label{phizz}
\Phi_{zz}(t,z)&\leq& -\delta_\varepsilon, \ \ \ \ \forall t\in[\bar{t},\bar{t}+\varepsilon), \ z\in \mathcal{O}_t^\varepsilon.
\enq
\end{enumerate}
\end{Lemma}
\textbf{Proof.} 1. Set
\beq\label{JJ}
\mathcal{J}(t,z;\hat{c},\hat{\pi})\;:=\; \widehat \E\left[\int_0^\infty e^{-K_{\lambda} s} \left(\frac{\hat{c}^p_s}{p}+\lambda f_\gamma (t+s,Z_s^{0,z,\hat{c},\hat{\pi}}\right)ds\right].
\enq
As the SDE for $Z$ is autonomous, we have
 \beq\label{VV}
\Phi(t,z)\;=\;\sup_{(\hat{c},\hat{\pi})\in\hat{\mathcal{A}}_0(z))} \mathcal{J}(t,z;\hat{c},\hat{\pi}).
 \enq
Recall that
 $f_\gamma(t,z) ={K_V}\ \E (z+J_t)^p$. Applying Dynkin's formula to ${K_V} (z+J_t)^p$, we see that  $f_{\gamma}(\cdot,z)$ is differentiable and
\beq\label{PP}
\left|\,\frac{\partial}{\partial t} f_\gamma(t,z)\,\right|\ \ \leq\ \  C_{J,p}   f_\gamma(t,z),
\enq
where
$
C_{J,p} = |b_J|p+\frac{1}{2}p(1-p)\sigma_J^2.
$
So we can differentiate  \eqref{JJ} with respect to $t$ and, using \eqref{PP}, we  get
$
\left|\,\frac{\partial}{\partial t}\,\mathcal{J}(t,x,y;c,\pi)\,\right|
\leq  C_{J,p}\ \Phi(t,z).
$
The latter estimate  is uniform in $(\hat{c},\hat{\pi})\in\hat{\mathcal{A}}_0(z)$, so from \eqref{VV} and the fact that
$$|\Phi(t+h,z)-\Phi(t,z)|\ \ \leq \ \ \sup_{(\hat{c},\hat{\pi})\in\hat{\mathcal{A}}_0(z)} |\mathcal{J}(t+h,z;\hat{c},\hat{\pi})-\mathcal{J}(t,z;\hat{c},\hat{\pi})|,$$
 we get the claim with
$N_\varepsilon =  C_{J,p} \cdot\sup_{D_\varepsilon(\bar{t},\bar{z})}  \Phi(t,z).$

2. Let $(t,z)\in D_\varepsilon(\bar{t},\bar{z})$.
  $\Phi(t,\cdot)$ is  strictly increasing for each $t$\,: this follows from the fact that it is concave, nondecreasing (it inherits these properties from $\widehat{V}$), and that  $\lim_{z\rightarrow+\infty}\Phi(t,z)=+\infty$. Hence, strict monotonicity and concavity yield
 the existence of the left and right derivatives $\Phi_z^-(t,z)$,  $\Phi_z^+(t,z)$, and the inequalities
$\Phi_z^-(t,z)\geq  \Phi_z^+(t,z)> 0.$
Then,  to show that $\Phi(t,\cdot)$ is differentiable at $z$, we need to prove that the first of the previous  inequality is actually an equality. Assume,  by contradiction, that  $\Phi_z^-(t,z)>  \Phi_z^+(t,z)$. Let $\delta>0$ and consider the function, defined for $ z_1\in (\bar{z}-\varepsilon, \bar{z}+\varepsilon), \ t_1\in[t,\bar{t}+\varepsilon)$,
 $$\varphi^\delta(t_1,z_1)\;:=\; \Phi(t,z)+\frac{\Phi_z^-(t,z)+ \Phi_z^-(t,z)}{2}(z_1-z)-\frac{1}{2\delta}(z_1-z)^2+(N_\varepsilon+\delta)(t_1-t).$$
 Due to  item 1, the function $\Phi-\varphi^\delta$ has a local maximum at $(t,z)$  in  $(\bar{z}-\varepsilon, \bar{z}+\varepsilon)\times [t,\bar{t}+\varepsilon)$. Therefore, the subsolution viscosity property at $(t,z)$  implies
 \begin{multline*}
- N_\varepsilon-\delta+ K_\lambda \Phi(t,z) - K_2 z \frac{\Phi_z^-(t,z)+ \Phi_z^-(t,z)}{2} - \lambda {K_V} f_\gamma(t,z)+\frac{K_\gamma^2}{2} z^2 \frac{1}{\delta}\\
-\sup_{\hat{c}\geq 0,\, {\hat{\pi}}\in\R } {H}^1_{cv}\left(\frac{\Phi_z^-(t,z)+ \Phi_z^+(t,z)}{2},-\frac{1}{\delta};\hat{c},{\hat{\pi}}\right)\;\leq \; 0,
 \end{multline*}
 where
$
H^1_{cv}(r,q;\hat{c},\hat{\pi})= \frac{\hat{c}^p}{p}-\hat{c}r
+ K_1\hat{\pi}r+\frac{1}{2}\sigma_L^2\hat{\pi}^2_2q.
$
 Letting $\delta\rightarrow 0$, we get a contradiction as $\frac{K_\gamma^2}{2} z^2 \frac{1}{\delta}\to+\infty$, {whereas the other terms are bounded uniformly in $\delta$.} So $\Phi(t,\cdot)$ is differentiable at each $z\in(\bar{z}-\varepsilon,\bar{z}+\varepsilon)$  for every $t\in [\bar t,\bar{t}+\varepsilon)$. The fact that  $\Phi(t,\cdot)\in C^1((\bar{z}-\varepsilon,\bar{z}+\varepsilon);\R)$, for every $t\in [\bar t,\bar{t}+\varepsilon)$, follows from concavity. Finally, let us show \eqref{phiz}. Let $\delta=\frac{\bar{z}-\varepsilon}{2}$. By concavity of $\Phi(t,\cdot)$, we have, for every $z\in(\bar{z}-\varepsilon,\bar{z}+\varepsilon)$ and $t\in [\bar t,\bar{t}+\varepsilon)$,
 \beq\label{bas}
 \frac{\Phi(t,\bar{z}-\varepsilon)-\Phi(t,\bar{z}-\varepsilon-\delta)}{\delta}\ \leq\  \Phi_z(t,z)\ \leq\  \frac{\Phi(t,\bar{z}+\varepsilon+\delta)-\Phi(t,\bar{z}+\varepsilon)}{\delta},
 \enq
 and, by concavity and strict monotonicity,
 $$\frac{\Phi(t,\bar{z}-\varepsilon)-\Phi(t,\bar{z}-\varepsilon-\delta)}{\delta}\ <\ +\infty, \ \ \ \ \frac{\Phi(t,\bar{z}+\varepsilon+\delta)-\Phi(t,\bar{z}+\varepsilon)}{\delta}
\ >\ 0.$$
Calling
 $M_\varepsilon :=  \sup_{t\in [\bar t,\bar{t}+\varepsilon)} \frac{\Phi(t,\bar{z}-\varepsilon)-\Phi(t,\bar{z}-\varepsilon-\delta)}{\delta}, \ m_\varepsilon\ :=\  \inf_{t\in [\bar t,\bar{t}+\varepsilon)}\frac{\Phi(t,\bar{z}+\varepsilon+\delta)-\Phi(t,\bar{z}+\varepsilon)}{\delta}, $
 by continuity of $\Phi$, we have $0<m_\varepsilon\leq M_\varepsilon<\infty$, so the claim follows by \eqref{bas}.

 3. Let  $(t,z)\in D_\varepsilon(\bar{t},\bar{z})$. The fact that there exists a set $\mathcal{O}_t^\varepsilon$ with full Lebesgue measure such that $\Phi(t,\cdot)$ is differentiable at the points of $\mathcal{O}_t^\varepsilon$ follows from concavity of $\Phi(t,\cdot)$ and Alexandrov's Theorem. Assume $z\in \mathcal{O}_t^\varepsilon$. Letting $\delta>0$ and $\delta_1>0$, consider the function defined, for $z_1\in (\bar{z}-\varepsilon, \bar{z}+\varepsilon), \ t_1\in[t,\bar{t}+\varepsilon)$, as
 $$\varphi^\delta(t_1,z_1)\;:=\; \Phi(t,z)+{\Phi_z}(t,z)(z_1-z)+\frac{1}{2}(\Phi_{zz}(t,z)-\delta) (z_1-z)^2-(N_\varepsilon+\delta_1)(t_1-t).$$
 Due to  item 1, the function $\Phi-\varphi^\delta$ has a local minimum at $(t,z)$ in  $(\bar{z}-\varepsilon, \bar{z}+\varepsilon)\times [t,\bar{t}+\varepsilon)$, for each $\delta>0$.
 Therefore, the supersolution viscosity property at $(t,z)$ and item 2 imply
 \begin{multline}\label{hhj}
N_\varepsilon+\delta_1+ K_\lambda \Phi(t,z) - K_2 z {m_\varepsilon} - \lambda {K_V}f_\gamma(t,z)-\frac{K_\gamma^2}{2} z^2(\Phi_{zz}(t,z)-\delta)\\-\tilde{U}(M_\varepsilon)+\frac{1}{2}\frac{K_1^2}{\sigma_L^2}\frac{m_\varepsilon^2}{\Phi_{zz}(t,z)-\delta} \ \geq \ 0.
 \end{multline}
Note that, given  $a_0, b_0>0$ and $c_0\in\R$, there exists $\alpha_0>0$ such that
\beq\label{ret}
a_0\xi-\frac{b_0}{\xi}\;\leq\; c_0,\ \ \xi\;\leq\; 0 \ \Longrightarrow \ \xi\;\leq\; - \alpha_0.
\enq
As $\Phi_{zz}\leq 0$, from \eqref{hhj} we see that \eqref{ret} holds for $\xi=\Phi_{zz}(t,z)-\delta$. So we get the existence of $\delta_\varepsilon>0$, independent of $(t,z)\in D_\varepsilon(\bar{t},\bar{z})$ and of $\delta$, such that
$
 \Phi_{zz}(t,z)\leq \delta - \delta_\varepsilon.
$
By arbitrariness of $\delta$ we get the claim.
\hfill\ep

\begin{Proposition}\label{prop:HJBv}
$\Phi$ is a viscosity solution in $D_\varepsilon(\bar{t},\bar{z})$ of
\beq\label{HJBphi3}
- \varphi_t + K_\lambda \varphi  -\! \lambda f_\gamma(t,z) {K_V} - \sup_{\hat{c}\in[0,{c_M}],\, {\hat{\pi}}\in [-{\pi_M},{\pi_M}]}{H}^0_{cv}(z, \varphi_z,\varphi_{zz};\hat{c},{\hat{\pi}})\  =\  0,
\enq
where
${c_M}:= (U')^{-1}(m_\varepsilon), \ {\pi_M} := \frac{|K_1| M_\varepsilon}{\sigma_L^2\delta_\varepsilon}.$
\end{Proposition}
\textbf{Proof.} The fact that $\Phi$ is a supersolution of \eqref{HJBphi3}  in $D_\varepsilon(\bar{t},\bar{z})$ follows from  the fact that it is a supersolution of \eqref{HJBphi*}, as the supremum is taken over a smaller set in \eqref{HJBphi3}. Let us show that it is a subsolution  in $D_\varepsilon(\bar{t},\bar{z})$. Take $(t, z)\in  D_\varepsilon(\bar{t},\bar{z})$ and let $\varphi\in C^{1,2}(D_\varepsilon(\bar{t},\bar{z});\R)$ be such that $\varphi(t,z)=\Phi(t,z)$ and   $ \varphi\geq \Phi$ in $D_\varepsilon(\bar{t},\bar{z})$. As $\Phi$ is differentiable with respect to $z$, it must be $\varphi_z(t,z)=\Phi_z(t,z)$.
If $\varphi_{zz}\leq -\delta_{\varepsilon}$, then
\beq\label{ghs1}
\sup_{\hat{c}\geq 0,\, {\hat{\pi}}\in\R } {H}^0_{cv}(z,{\varphi}_z,{\varphi}_{zz};\hat{c},{\hat{\pi}})&=& \sup_{\hat{c}\in[0,{c_M}],\ {\hat{\pi}}\in [-{\pi_M},{\pi_M}]}{H}^0_{cv}(z, {\varphi}_z,{\varphi}_{zz};\hat{c},{\hat{\pi}}),
\enq
so we have the desired subsolution inequality. Otherwise, assume $\varphi_{zz}(t,z)>-\delta_\varepsilon$ and
consider the function $\tilde{\varphi}$ defined, for $ z_1\in (\bar{z}-\varepsilon, \bar{z}+\varepsilon), \ t_1\in[t,\bar{t}+\varepsilon)$, as
\beq \label{vvb}
\tilde{\varphi}(t_1,z_1)\;:=  \;  \varphi(t_1,z)+\Phi_z(t_1,z)(z_1-z)-\frac{1}{2}\delta_\varepsilon(z_1-z)^2.
\enq
We have
\beq\label{pkjs}
\tilde{\varphi}(t_1,z)\;\geq\;\varphi(t_1,z)\;\geq\; \Phi(t_1,z), \ \  \ \  \forall \, t_1\in[t,\bar{t}+\varepsilon).
\enq
Fix $t_1\in[t,\bar{t}+\varepsilon)$.
Consider, for $z_1\in  (\bar{z}-\varepsilon, \bar{z}+\varepsilon)$,  the Dini derivative of $\Phi_z$ at $z_1$, i.e. $
D^+_z\Phi_{z}(t_1,z_1)\;:=\; \limsup_{h\rightarrow 0} \frac{\Phi_z(t_1,z_1+h)-\Phi_z(t_1,z_1)}{h}.
$
As $\Phi(t_1,\cdot)$ is concave, we have
\beq\label{dini}
D^+_z\Phi_{z}(t_1,z_1)\;\leq\; 0, \ \ \  \forall \ z_1\in  (\bar{z}-\varepsilon, \bar{z}+\varepsilon).
\enq
Moreover, by  Lemma \ref{lemma:sub}(3), we have
\beq\label{dini2}
D^+_z\Phi_{z}(t_1,z_1)\;\leq \;-\delta_\varepsilon, \ \ \  \forall \ z_1\in  \mathcal{O}_{t_1}^\varepsilon.
\enq
From \eqref{dini}-\eqref{dini2}, from the fact that $\mathcal{O}_{t_1}^\varepsilon$ has full measure, and from Lemma 3.3 in \cite{FGG11}, we get,
by integrating twice \eqref{dini2},
\beq\label{dini3}
\Phi(t_1,z_1)\;\leq\; \Phi(t_1,z)+\Phi_z(t_1,z)(z_1-z)-\frac{1}{2}\delta_\varepsilon(z_1-z)^2.
\enq
Combining \eqref{dini3} with \eqref{vvb}-\eqref{pkjs}, we get
$
\tilde{\varphi}(t,z)=\Phi(t,z)$ and   $\tilde{\varphi}\geq \Phi \ \mbox{in} \  (\bar{z}-\varepsilon, \bar{z}+\varepsilon)\times [t,\bar{t}+\varepsilon).
$
As $\Phi$ is a viscosity subsolution of \eqref{HJBphi*}, we have
\beqs
- \tilde{\varphi}_t + K_\lambda \tilde{\varphi}  - \lambda{K_V} f_\gamma(t,z)-\sup_{\hat{c}\geq 0,\, {\hat{\pi}}\in\R } {H}^0_{cv}(z, \tilde{\varphi}_z,\tilde{\varphi}_{zz};\hat{c},{\hat{\pi}})\ \leq\ 0.
\enqs
On the other hand,
\beqs\label{ghs}
\sup_{\hat{c}\geq 0,\, {\hat{\pi}}\in\R } {H}^0_{cv}(z,\tilde{\varphi}_z,\tilde{\varphi}_{zz};\hat{c},{\hat{\pi}})&=& \sup_{\hat{c}\in[0,{c_M}],\ {\hat{\pi}}\in [-{\pi_M},{\pi_M}]}{H}^0_{cv}(z, \tilde{\varphi}_z,\tilde{\varphi}_{zz};\hat{c},{\hat{\pi}}),
\enqs
so  also
\beq\label{HJBphi4}
- \tilde{\varphi}_t + K_\lambda \tilde{\varphi} - \lambda {K_V}f_\gamma(t,z)-\sup_{\hat{c}\in[0,{c_M}],\ {\hat{\pi}}\in [-{\pi_M},{\pi_M}]} {H}^0_{cv}(z,\tilde{\varphi}_z,\tilde{\varphi}_{zz};\hat{c},{\hat{\pi}})\;\leq\;0.
\enq
Noting that
\beq\label{fga}
\varphi(t,z)\;=\;\tilde{\varphi}(t,z), \ \ \ \varphi_z(t,z)\;=\;\tilde{\varphi}_z(t,z),\ \ \ \ \varphi_{zz}(t,z)\;>\;-\delta_\varepsilon\; =\; \tilde{\varphi}_{zz}(t,z),
\enq
and taking into account that ${H}^0_{cv}$ is nondecreasing in the last argument, combining \eqref{HJBphi4} and \eqref{fga}, we get the subsolution inequality for $\varphi$. \hfill\ep

\begin{Lemma} \label{lemDiffVisc}
Let $a<b$ and let $F$ : $[0,T) \times (a,b) \times \R^3$ $\to$ $\R$, $(t,x,r,p,q) \mapsto F(t,x,r,p,q)$, be continuous, continuously differentiable in $(x,r,p,q)$, and proper degenerate elliptic (i.e. nondecreasing in $r$ and nonincreasing in $q$).
Let $u\in C^{1,2}([0,T)\times (a,b);\R)$ be a classical  solution in   $[0,T) \times (a,b)$ to
\beq \label{eqPDE1}
u_t + F(t,x,u,u_x,u_{xx}) &=& 0.
\enq
Then the space derivative $v:=u_x$ is a viscosity solution in $[0,T) \times (a,b)$ to
\beq \label{eqPDE2}
v_t + \nabla F(t,x,u(t,x), v, v_x) \cdot (1,v,v_x,v_{xx}) = 0,
\enq
where $\nabla F$ $=$ $(F_x,F_r,F_p,F_q)$.
\end{Lemma}

\textbf{Proof.}
For $x\in (a,b)$ and sufficiently small $h>0$, define $u^h(t,x) := u(t,x+h)$ and $v^h := \frac{u^h-u}{h}$. Then, due to continuous differentiability of $u$, we have   $v^h \to v$ locally uniformly in $[0,T)\times (a,b)$, when $h\rightarrow 0^+$. Furthermore, as $u$ is a solution to \reff{eqPDE1} and using the differentiability of $F$, we see that
\beqs
v^h_t &=& \frac{1}{h}\left( F(t,x,u^h,u^h_x,u^h_{xx}) - F(t,x,u,u_x,u_{xx})\right) \\
&=& \left( \nabla F(t,x,u,v^h,v^h_x) +  {\cal E}^h(t,x) \right)\cdot (1,v^h,v^{h}_x,v^h_{xx}),
\enqs
where
{\small
\beqs
 &&{\cal E}^h(t,x):= \\
& &\!\!\!\!\!\!\!\!\int_0^1\!\! \nabla F(t,x+hs,(1-s) u^h(t,x) + su(t,x),(1-s) u^h_x(t,x) + su_x(t,x),(1-s) u^h_{xx}(t,x) + su_{xx}(t,x)) ds \\
&&- F(t,x,u(t,x),v^h(t,x),v^h_x(t,x)).
\enqs}
By continuity of $F$ and of $u$, $u_x$, $u_{xx}$, and by the fact that $v^h$ (respectively, $v^h_x$) goes to $u_x$ (respectively, to $u_{xx}$), as $h\rightarrow 0^+$,
we see that  ${\cal E}^h\rightarrow 0^+$, as $h$ $\to$ $0^+$, locally uniformly in $[0,T)\times (a,b)$.
Applying the  stability result for viscosity solutions (see, e.g.,  \cite[Prop.\,5.9, Ch\,.4]{YZ}), we  get that $v$ is a viscosity solution to  \reff{eqPDE2}.
\hfill\ep

\subsection{A result by Kryolv on existence of  classical solutions to fully nonlinear parabolic equations}
\begin{Theorem}\label{teo:krylov}
Let $\Theta$ be an index set and let $Q:= (0,T) \times \mathcal{O}$, with $\mathcal{O}\subset \R^{N}$ open.
Let
$$a=(a^{i,j})_{i,j=1,...,N}: \Theta \times (0,T)\times \mathcal{O} \times \R \times \R^N\rightarrow \R^{N \times N},
$$
$$b : \Theta \times (0,T)\times \mathcal{O} \times \R \times \R^N\rightarrow \R,
$$
  and call  $(\theta, t, x, r, q)$ their formal arguments. Assume the following conditions.
\begin{enumerate}
\item For every $\theta\in \Theta$, the functions $a,b$ are continuously differentiable with respect to $(t,x,r,q)$, and, for every $(\theta,t)\in \Theta\times (0,T)$, they are twice continuously differentiable with respect to $(x,r,q)$.
\item The first derivatives of $a,b$ with respect to $t$ and the second derivatives of $a,b$ with respect to $(x,r,q)$ are bounded in every set of the form $$S_M:= \{(\theta,t,x,r,q) \in \Theta \times (0,T)\times \mathcal{O} \times \R \times \R^N\  : \ \theta\in \Theta, (t,x)\in Q, r+|q|\leq M\}.$$
\item The function  $a$ satisfies a uniform ellipticity condition\,: for some constants $\Lambda\geq \varepsilon>0$,
\begin{equation*}
\varepsilon |\xi|^2 \leq \sum_{i,j} a^{i,j} \xi_i \xi_j \leq \Lambda |\xi|^2, \ \ \ \forall (\theta, t, x, r, q)\in\Theta \times (0,T)\times \mathcal{O} \times \R \times \R^N, \ \forall \xi\in\R^N.
\end{equation*}
\item
There exists a continuous function $h$ such that, for every $(\theta,t,x,r,q) \in \Theta \times (0,T)\times \mathcal{O} \times \R \times \R^N$,
$$|D_qa^{i,j}| (1+|q|)+|D_r a^{i,j}|+|D_xa^{i,j}|(1+|q|)^{-1}\leq h(r), \ \ \ \forall i,j=1,...,N,$$
$$
|D_qb| (1+|q|)+|b|+ |D_r b|+|D_x b|(1+|q|)^{-1} \leq h(r)(1+|q|^2).
$$
\item There exist constants $\delta_0>0 $ and  $M_0>0$ such that
$$
b(\theta, t,x, -M_0,0)\geq \delta_0, \ \ \ \ b(\theta, t,x, M_0,0)\leq -\delta_0,  \ \ \ \forall (\theta,t,x)\in \Theta\times Q.$$
%
\end{enumerate}
Then, for each $\phi$ in $C(\bar Q)$, there exists a unique $u \in C^{1,2}(Q) \cap C(\bar Q)$ solution to
\begin{equation*}
-\partial_t u - \sup_{\theta \in \Theta} \left\{ \operatorname{Tr}\left(a(\theta,t,x,u,Du) D^2u\right) + b(\theta,t,x,u,Du) \right\} =0\;\; \mbox{ in } Q,
\end{equation*}
with Dirichlet boundary condition
$
u = \phi$ on   $\mathcal{P} Q  := \{T\} \times \mathcal{O} \cup (0,T) \times \partial\mathcal{O}.
$
\end{Theorem}

\textbf{Proof.}
See \cite[Th.\,3, Sec.\,6.4, p.\,301]{K87}. The conditions are those in Example 8, Section 6.1, p.\,279, of the same book.
\hfill$\square$
\vspace{.5cm}\noindent{\small{\textbf{Acknowledgements.} The authors thank the Editor, the Associate Editor and two anonymous Referees for their careful reviews and their useful suggestions. {The authors are sincerely grateful to Silvia Faggian for her kind help for the editing work and Claudio Tebaldi for valuable comments.} The authors also thank, for useful comments, the participants to the seminars where this paper has been presented, in Berne, Florence, Levico Terme, Pise, Bruxelles, Berlin, Manchester.}


\begin{thebibliography}{}

{\small{

\bibitem{APW} Ang A.,  Papanikolaou D., and M. Westerfield (2014) : ``Portfolio Choice with Illiquid Assets", \textit{Management Science}, Vol. 60, No. 11.

\bibitem{BW96} Beaudry P. and  E. van Wincoop (1996) : ``The Intertemporal Elasticity of Substitution: An Exploration Using a US Panel of State Data," Economica, London School of Economics and Political Science, Vol. 63, pp. 495-512.


\bibitem{BL10} Bayraktar E. and B. Ludkovski (2011) : ``Optimal Trade Execution in Illiquid Markets", \textit{Mathematical Finance}, Vol.\,21(4), pp.\,681--701.

\bibitem{BS13} Bayraktar E. and M. S\^irbu (2013) : ``Stochastic Perron's method for Hamilton-Jacobi-Bellman equations'', \textit{SIAM Journal on Control and Optimization} , 51(6), 4274-4294, 2013.

\bibitem{bayzho} Bayraktar E. and Z. Zhou (2013) : ``On controller-stopper problems with jumps and their applications to indifference pricing of American options", SIAM Journal on Financial Mathematics, Vol. 5, No. 1, pp. 20-49.





\bibitem{CIL92} Crandall M., Ishii H., and P.L. Lions (1992) : ``User's Guide to Viscosity Solutions of Second Order Partial Differential Equations", {\it Bull. Amer. Math. Soc.}, {Vol. 27}, 1-67.






\bibitem{CGPT} Cretarola A., Gozzi F., Pham H., and  P. Tankov (2011) : ``Optimal consumption policies in illiquid markets", {\it
Finance and Stochastics},  Vol. 15, No. 1, pp. 85-115.


\bibitem{DFG} Di Giacinto M., Federico S., and F. Gozzi  (2011) : ``Pension funds with a minimum guarantee: a stochastic control approach", {\it Finance and Stochastics}, Vol. 15, No.\,2, 297-342.



\bibitem{FGtec} Federico S. and P. Gassiat (2014) : ``Viscosity characterization of the value function of an investment-consumption problem in presence of illiquid assets".  \textit{Journal of Optimization: Theory and Applications,} Vol. 160, No. 3, pp. 966--991.
\bibitem{FGGdual} Federico S., Gassiat P., and F. Gozzi (2012) : ``Utility maximization  with current  utility on the  wealth: regularity of solutions to the HJB equation". Forthcoming on \emph{Finance and Stochastics}. Preprint Arxiv.

\bibitem{FGG11} Federico S., Goldys B., and F. Gozzi (2011) : ``HJB equations for the optimal control of differential equations with delays and state constraints: verification and optimal feedbacks", \emph{SIAM - Journal on Control and Optimization}, Vol.\,49, No.\,6, pp.\,2378--2414.

\bibitem{FS06} Fleming W. H. and H. M. Soner (2006) : \textit{Controlled Markov Processes and Viscosity Solutions}, Springer-Verlag.

\bibitem{gasphd} Gassiat P. (2011) :  ``Mod\'elisation du risque de liquidit\'e et m\'ethodes de quantification appliqu\'ees au contr\^ole stochastique s\'equentiel", Phd thesis of University Paris Diderot, available at \texttt{http://tel.archives-ouvertes.fr/tel-00651357/fr/}

\bibitem{gasgozpha11} Gassiat P., Gozzi F., and H. Pham (2014): ``Investment/consumption problem in illiquid markets with regime-switching",
\textit{SIAM Journal on Control and Optimization}, Vol. 52, No. 3, pp. 1761-1786.

\bibitem{Juu01} Juutinen, P. (2001) : ``On the Definition of Viscosity Solutions for Parabolic Equations", \textit{Proc. Amer. Math. Soc.}, Vol. {129}, No. 10, pp. 2907-2911.






\bibitem{Kab} Kabanov Y., and M. Safarian (2009) : ``Markets with Transaction Costs: Mathematical Theory", Springer-Verlag.

\bibitem{Kallenberg} Kallenberg O. : \textit{Foundations of modern probability}, Second ed., Probability and its
Applications, Springer-Verlag, New York, 2002.

\bibitem{KS} Karatzas I. and S.E. Shreve (1988) : \textit{Brownian motion and stochastic calculus}, Springer-Verlag, New York.




\bibitem{K87} Krylov N.V. (1987) : \emph{Nonlinear elliptic and parabolic equations of the second order}, D. Reidel Publishing Company.



\bibitem{L05} Longstaff F. (2005) : ``Asset pricing in markets with illiquid assets", \textit{American Economic Review}, Vol. 99, No. 4, pp. 1119--1144.

\bibitem{mat06} Matsumoto K. (2006) : ``Optimal portfolio of low liquid assets with a log-utility function", {\it Finance and Stochastics},
{Vol. 10},  No. 1, pp. 121--145.

{\bibitem{pha10}  Pham H.,  {Stochastic control under progressive enlargment of filtrations and applications to multiple defaults risk management}, {Stoch. Proc. and their Appl.}, 120, 1795-1820 (2010)}


 \bibitem{phatan08} Pham H. and P. Tankov (2008) : ``A model of optimal consumption under liquidity risk with random trading times",
 {\it Mathematical Finance}, {Vol. 18}, No. 4,  pp. 613--627.

\bibitem{phatan09} Pham H. and P. Tankov (2009) : ``A coupled system of integrodifferential equations arising in liquidity risk model", \textit{Applied Mathematics and Optimization}, Vol. 59, No. 2, pp. 147--173.

\bibitem{RZ02} Rogers C. and O. Zane (2002) : ``A simple model of liquidity effects", in \textit{Advances in
Finance and Stochastics: Essays in Honour of Dieter Sondermann}, Eds. K. Sandmann
and P. Schoenbucher, pp. 161--176.

\bibitem{ST04} Schwartz E. and C. Tebaldi (2006) : ``Illiquid assets and optimal portfolio choice",
NBER Working Paper No. w12633.



\bibitem{YZ} Yong J., and X.Y. Zhou (1999) : ``Stochastic Controls: Hamiltonian Systems and HJB Equations",  Springer Verlag, New York.
%

}}
\end{thebibliography}
\end{document}